\documentclass[preprint,a4paper,3p,10pt,onecolumn]{elsarticle}
\usepackage{color}
  \usepackage{newlfont}
  \usepackage{graphicx}
  \usepackage{epstopdf}
  \usepackage{amssymb}
  \usepackage{amsmath}
  \usepackage{bm}
  \usepackage{verbatim}
  \usepackage[latin1]{inputenc}
  \usepackage{bbm}
  \usepackage{mathptmx}
  \usepackage{lineno,hyperref}
  \modulolinenumbers[5]
  
\journal{Physica A: Statistical Mechanics and its Applications}

\newcommand{\be}{\begin{eqnarray}}
\newcommand{\ee}{\end{eqnarray}}
\begin{document}

\begin{frontmatter}

\title{Generalized discord, entanglement, Einstein-Podolsky-Rosen steering, and Bell nonlocality in two-qubit systems under (non-)Markovian channels: Hierarchy of quantum resources and chronology of deaths and births}


\author[UFPR]{A. C. S. Costa}
\ead{ana.sprotte@gmail.com}
\author[UFPR]{M. W. Beims}
\author[UFPR]{R. M. Angelo}
\address[UFPR]{Departamento de F\'isica, Universidade Federal do Paran\'a, 81531-980, Curitiba, Brazil}


\begin{abstract}
Generalized quantum discord $(D_q)$, Einstein-Podolsky-Rosen steering $(S)$,  entanglement $(E)$, and Bell nonlocality ($N$), are logically distinct quantifiers of quantum correlations. All these measures capture nonclassical aspects of quantum states and play some role as resources in quantum information processing. In this work, we look for the hierarchy satisfied by these quantum correlation witnesses for a class of two-qubit states. We show that $N \triangleright S\triangleright E\triangleright D_q$, meaning that nonlocality implies steering, which in turn implies entanglement, which then implies $q$-discord. For the quantum states under concern, we show that the invariance of this hierarchy under noisy quantum channels directly implies a death chronology. Additionally, we have found that sudden death of all quantum resources except discord is absent only for a subset of states of measure zero. At last, we provide an illustration of another consequence of the aforementioned hierarchy, namely, the existence of a sudden birth chronology under non-Markovian channels.
\end{abstract}




\end{frontmatter}


\section{Introduction}

{\em Correlations} are essential elements in physical theories. Every observation made in the laboratory is ultimately interpreted through such conceptual connections between the prepared system and read outcome.  Hence, the information one can obtain from nature is somehow determined by how much the system of interest can get correlated, via physical interactions, with the apparatus. Not surprisingly, therefore, correlations occupy a prominent place in information science. As far as the ``quantum versus classical'' dilemma is concerned, the question naturally arises whether the quantumness of correlations plays some distinctive role. It turns out that to date there is a significant amount of work showing that the answer is to be given in the positive. In particular, quantum correlations have shown to be at the core of fundamental physical arenas, from foundational phenomena, such as nonlocality~\cite{epr,bell,popescu94}, decoherence~\cite{zurek03}, and emergent reality~\cite{angelo15,bill15}, to the promising field of quantum information and computation~\cite{vincenzo,chuang,delgado02,wolf10}. 

{\em Entanglement} emerges in this context as a class of correlations that cannot be prepared by local operations and classical communication~\cite{h409}. Believed to be the main responsible for the speed-up in quantum computation and quantum communication, entanglement is essential for the realization of many subtle protocols, as for instance quantum teleportation~\cite{wootters93,furusawa13,wallraff13}. As far as mixed states are concerned, however, it is {\em quantum discord}~\cite{zurek01,vedral01}---a quantifier of measurement-induced disturbance---that appears as a more fundamental measure of quantumness, as it can assume nonzero values even for separable states. Like entanglement, quantum discord has been shown to be a physical resource for quantum protocols~\cite{walther12,lam12,brodutch13}, with the advantage of being less fragile to noisy channels~\cite{boas09,merali11} (see also Refs.~\cite{celeri11,vedral12} for recent reviews about quantum discord). 

It is to be joined to this framework the notion of {\em steering}, a term coined in 1935 by Schr\"{o}dinger~\cite{schrodinger35,schrodinger36} within the context of the Einstein-Podolsky-Rosen (EPR) paradox to name Alice's ability in affecting Bob's state through her choice of measurement basis. Steering has been formalized in terms of a quantum information task involving bipartite states and measurement settings~\cite{doherty07,doherty07_2,reid09}, in which case the existence of entanglement is necessary but not sufficient. Recently, a closed formula for general two-qubit systems has been proposed~\cite{costa16} for two- and three-measurements per site. Steerability of quantum states had been used for tasks involving randomness generation~\cite{law14}, subchannel discrimination~\cite{piani15}, quantum information processing~\cite{branciard11}, and one-sided device-independent processing in quantum key distribution (QKD)~\cite{branciard12}, within which context a resource theory of steering has recently been formulated ~\cite{gallego15}. Experimentally, steerability have been reported for Bell-local entangled states~\cite{saunders10}, entangled Gaussian modes of light~\cite{handchen12}, and also in loophole-free experiments~\cite{wittmann12,smith12,bennet12}. 

Finally, there is {\em Bell nonlocality}~\cite{brunner14}, a feature of those quantum states whose correlations cannot be explained in terms of local hidden variables. Revealed by violations of Bell inequalities~\cite{bell,chsh}, nonlocality has proven useful for QKD~\cite{scarani07,masanes06,kent05}, randomness generation~\cite{monroe10}, and quantum communication complexity~\cite{zeilinger04}. An inequality involving more than two-measurements per site, can be found in the Ref.~\cite{collins04}.

Despite the indisputable relevance of the aforementioned witnesses of quantumness, it is fair to say that to the date it is still not completely clear what is the essential connection among them, if any. Although these witnesses are logically distinct, all of them agree in diagnosing whether a pure bipartite state is quantum correlated. On the other hand, for mixed states the situation is quite different: There exist discordant states with no entanglement, entangled states with no steering, and steerable states with no nonlocality. On the formal side, besides offering a rigorous definition for steering, Wiseman {\em et al}~\cite{doherty07} proved that steerable states are a strict subset of the entangled states, and a strict superset of the states that can exhibit Bell nonlocality. These findings establish a formal hierarchy among entanglement, steering, and Bell nonlocality specifically for the case of an infinite number of measurements, a clearly impracticable situation. As far as realistic situations are considered, for which only a finite number of settings are allowed and decoherence is generally inevitable, it is not clear whether this hierarchy will maintain, to which extent, and what its consequences are. 

This paper is devoted to discuss the above mentioned issues. We show that resources such as discord, entanglement, steering, and Bell nonlocality obey a given hierarchy and, therefore, are fundamentally different. Our study focus on the analysis of two-qubit X states and their dynamical evolution under general noisy quantum channels. In Sec.~\ref{hierarchy1}, we introduce our notion of hierarchy in terms of resource diagrams. Sec.~\ref{measures} introduces the measures of quantum correlations we are interested in, namely, $q$-discord, entanglement, steering and Bell nonlocality. Section~\ref{results} is reserved to our main results, involving hierarchy and robustness under noisy channels, and Sec.~\ref{conclusion} closes the paper with our final remarks.

\section{Hierarchy and chronology in $Q_2\times Q_1$ diagrams \label{hierarchy1}}

\begin{figure}[htb]
\centerline{\includegraphics[scale=0.15]{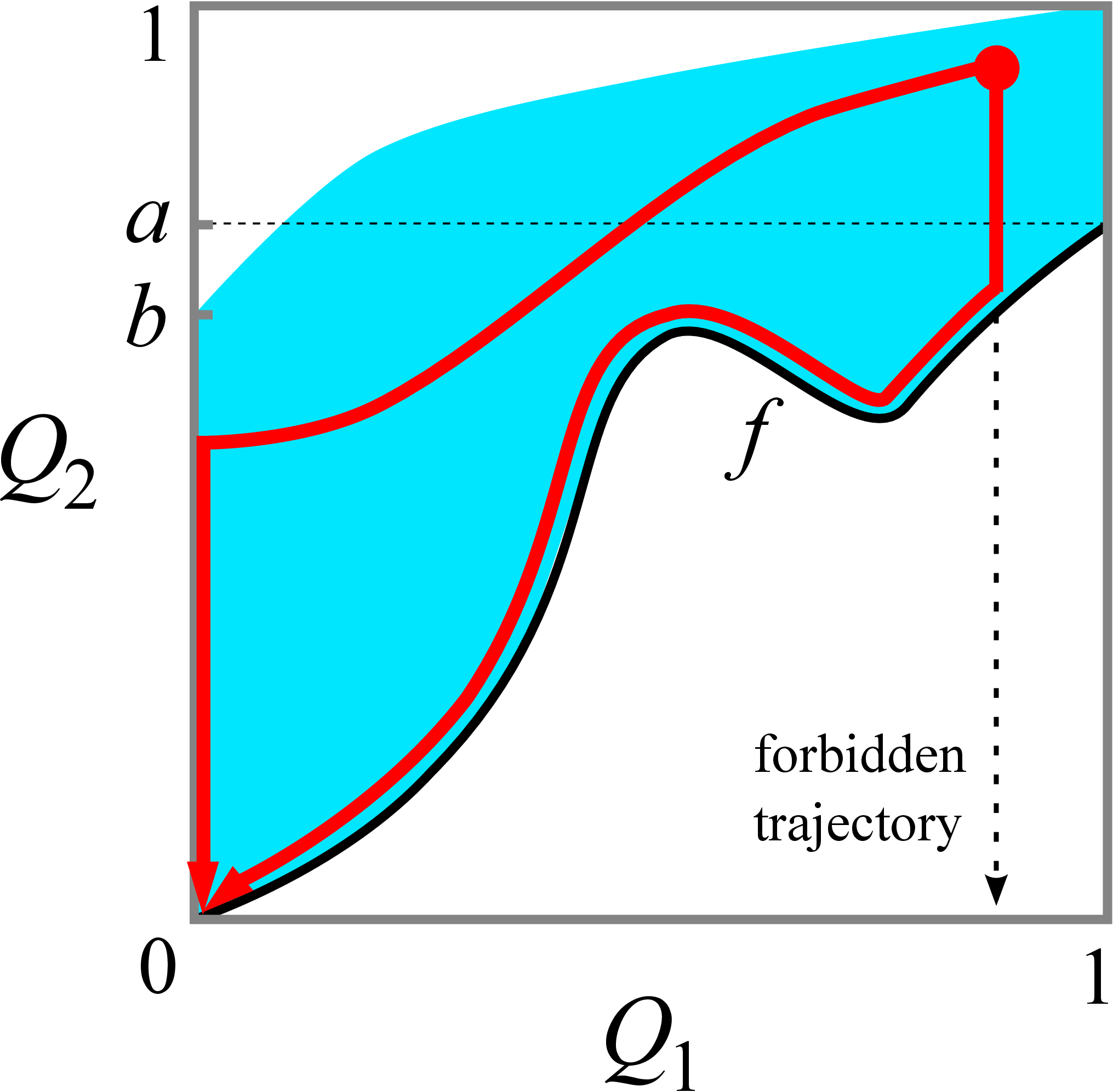}}
\caption{Hierarchy $Q_1 \triangleright  Q_2$ defined by the border $f$ (solid black line) for a given class of states.}
\label{fig1}
\end{figure}

Consider two different measures of quantum resources, $Q_1$ and $Q_2$, normalized in the real interval $[0,1]$. $(Q_1,Q_2)$ can assume any two of the measures under concern (discord, entanglement, steering, and nonlocality). Take a class of states $\rho_{\lambda}$, parametrized by a set $\lambda$ of parameters. Each state $\rho_{\lambda}$ corresponds to a point $(Q_1(\rho_{\lambda}),Q_2(\rho_{\lambda}))$ in a {\em resource diagram} $Q_2\times Q_1$. Typically, physical states are such that $\lambda$ is dense, so that the set of points in the diagram forms a region that is compact and connected (see the shaded area in Fig.~\ref{fig1}). We are now ready to introduce our notion of {\em hierarchy} between two measures of quantumness in terms of the diagram $Q_2\times Q_1$ depicted in Fig.~\ref{fig1}.

\vskip0.2cm
\noindent {\bf Definition 1} (Hierarchy).---{\em If for a given class of states $\rho_{\lambda}$
we can define a border $f$ such that $Q_2(\rho_{\lambda})\geqslant f(Q_1(\rho_{\lambda}))$, with $f(x)$ a non-negative continuous real function defined in the domain $x\in[0,1]$, such that $f(x)=0$ if and only if $x=0$ and $f(x)>0$ for $x>0$, then we say that a hierarchy denoted by $Q_1\triangleright Q_2$ holds, meaning that the existence of $Q_1$ for this class of states necessarily implies the existence of $Q_2$, while the converse is not true.}

\vskip0.2cm
\noindent {\bf Remark 1}.---{\em Let $f(x)$ be the tightest possible border, i.e., the one that limits the shaded region inferiorly. It follows that $\varepsilon f(x)$, with $\varepsilon <1$, is a border as well. Then, if $f$ exists, there will exist infinite borders. In addition, if $f$ exists, it will be possible to find a linear border  $f_{lin}=\epsilon x$ such that $f_{lin}\leqslant f$, equality for $x=0$. Therefore, it is enough to choose $\epsilon=\min_x \frac{f(x)}{x}$, $\forall\, x\in(0,1]$. On the other hand, choosing an arbitrarily small $\epsilon>0$ implies that the notion of hierarchy demands the exclusion of the line $(Q_1,0)$, with $Q_1\in(0,1]$.}
\vskip0.2cm

The notion defined above allows us to conclude, for a given class of states, that if $Q_1\triangleright Q_2$, then any effort to experimentally detect $Q_2$ can be reduced to detections of $Q_1$, as by the hierarchy it follows that the latter resource implies the former. This concept is specially useful in situations involving dynamics within a restrict space. To appreciate this point, consider a dynamical process $\Phi_t$, unitary or not, which maps a set of states onto itself, i.e., $\Phi_t(\rho_{\lambda(0)})=\rho_{\lambda(t)}$. In this case, any point in the diagram $Q_2\times Q_1$ will occupy another point in the same shaded area. As a consequence, during the dynamics the underlying hierarchy will always be respected. If the system is subjected to a Markovian noisy channel, in which all quantum correlations are expected to vanish in an irreversible way, the dynamics of any point in the diagram can be represented by a {\em trajectory} towards the {\em attractor} $(0,0)$. In particular, only two types of trajectories can be found: (i) either both resources $Q_1$ and $Q_2$ vanish simultaneously (lower thick red line in Fig.~\ref{fig1}) or (ii) $Q_1$ vanishes before $Q_2$ (upper trajectory), so as to obey the hierarchy $Q_1\triangleright Q_2$. On the other hand, for a non-Markovian process recurrences may occur, so that the trajectories may eventually resurge from $(0,0)$ and then posteriorly return to this attractor. 

The abrupt change in the upper trajectory clearly signalizes a {\em sudden death} of $Q_1$ whereas $Q_2> 0$. Let $t_{Q_1}$ be the instant at which the sudden death of $Q_1$ occurs. Later on, when the trajectory starts to follow the line $(0,Q_2>0)$, one can have either the sudden death of $Q_2$, say at a instant $t_{Q_2}$, or its asymptotic vanishing ($t_{Q_2}\to \infty$). In either case, the trajectory will reach attractor $(0,0)$. We thus come to another contribution of this work: the existence of a hierarchy $Q_1\triangleright Q_2$ necessarily implies a {\em death chronology} because the dynamics has to be such that $t_{Q_2}\geqslant t_{Q_1}$. This means that $Q_2$ is more robust than $Q_1$ to the lossy channel in question. If such a hierarchy is indeed universal for resources $Q_1$ and $Q_2$, then we should never find a vertical trajectory like in the $Q_2\times Q_1$ diagram. For non-Markovian processes, sudden birth may also occur, as we will show later.

A further important point is the following. Consider other genuine measures of quantum resources, $Q_k'$ ($k=1,2$), that can be written in the form $Q_k'=g_k(Q_k)$. Being $Q_k'$ genuine and normalized by assumption, we must demand that $g_k(x)=0$ if and only if $x=0$ and $g_k(x)>0$ for $x>0$. This is necessary in order to ensure that $Q_k'$ and $Q_k$ agree about the existence or absence of the pertinent resource. For concreteness, one may take as an example the link between entanglement of formation $Q_1'=E_{of}$~\cite{bennett96} and concurrence $Q_1=C$~\cite{wootters98}; an analytic monotonically increasing function $g$ is known to exist such that $E_{of}=g(C)$. Now we can prove the following result.
\vskip2mm

\noindent {\bf Lemma 1}.---{\em Given the hierarchy $Q_1(\rho_{\lambda})\triangleright Q_2(\rho_{\lambda})$ for a class of states $\rho_{\lambda}$ and functions $g_k(x)$ in the domain $x\in [0,1]$ such that $g_k(x)=0$ if and only if $x=0$ and $g_k(x)>g_k(0)$ for $x>0$, then $g_1(Q_1(\rho_{\lambda}))\triangleright g_2(Q_2(\rho_{\lambda}))$.}
\vskip2mm

\noindent\underline{Proof}: By Remark 1, the hypothesis $Q_1(\rho_{\lambda})\triangleright Q_2(\rho_{\lambda})$ implies the absence of the line $(Q_1,0)$, $\forall \,\, Q_1\in (0,1]$ in the $Q_2\times Q_1$ diagram. It readily follows from the properties of $g$ that the line $(Q_1',0)$  is excluded from the diagram $Q_2'\times Q_1'$. Therefore, the hierarchy $g_1(Q_1(\rho_{\lambda}))\triangleright g_2(Q_2(\rho_{\lambda}))$ holds. \hfill{$\blacksquare$}
\vskip2mm

\noindent This result is relevant because it ensures that we do not need to compare several measures of entanglement with several measures of nonlocality, and similarly for the other resources. It is enough to compare only a genuine pair. 

There are two other interesting aspects that can be revealed by resource diagrams (see Fig.~\ref{fig1}). First, if $a<1$ the situation is such that there are states for which $Q_1$ and $Q_2$ do not reach their maximal values simultaneously. As far as entanglement and Bell nonlocality are concerned, this instance has been referred to in the literature as an {\em anomaly}~\cite{methot07}. Second, if $a=1$ and $b=0$, the measures $Q_1$ and $Q_2$ agree in identifying states that have no correlation and those which are maximally correlated. What remains in this case is just the issue of ordering, which is well-known for entanglement measures~\cite{eisert99}. This motivates the notion of {\em equivalence}.

\vskip0.2cm
\noindent {\bf Definition 2} (Equivalence).---{\em If for states $\rho_{\lambda}$ the lines $(Q_1>0,0)$ and $(0,Q_2>0)$ are absent from the $Q_2\times Q_1$ diagram and, in addition, $Q_1$ reaches its extremal values (0 or 1) only when $Q_2$ does, then we say that an equivalence denoted by $Q_2\equiv Q_1$ holds for these measures for the the class of states in question.}
\vskip0.2cm

This means that in such a scenario we cannot refer to $Q_1$ and $Q_2$ as quantifiers of distinct resources.

In this section, we have defined notions such as hierarchy, death chronology, trajectory in the $Q_2Q_1$ space, anomaly, and equivalence within the context of a $Q_2\times Q_1$ diagram. Next we introduce the resource measures whose hierarchy we are interested in assessing.

\section{Quantum correlation measures \label{measures}}

It is consensual in the scientific community that discord, entanglement, steering, and Bell nonlocality are concepts associated with quantum correlations. However, since they are logically distinct in their essence, it seems inescapable to conclude that they either capture distinct aspects of the quantum correlations contained in a given state or reveal quantum correlations of distinct natures. Currently, the dominant view seems to regard these concepts as distinct resources in certain protocols of quantum information and quantum communication. Regardless of the view one may adopt, a relative characterization of such measures can provide some insight on the most fragile element defining the quantum correlations. In this section we introduce the measures we are concerned with.

\subsection{Generalized quantum discord}

When a projective measurement, defined by a set $\{\Pi_b\}$ of projectors acting on $\mathcal{H}_B$, is performed locally on a subsystem $B$, the global state $\rho\in \mathcal{H}_A\otimes\mathcal{H}_B$ changes. {\em Discord} is a quantifier that captures this measurement-induced disturbance. Originally, discord was introduced in entropic form as the difference between two classically equivalent versions of the mutual information~\cite{zurek01,vedral01}. Subsequently, a geometric version was derived in terms of the Schmidt norm~\cite{vedral10}. Recently, these two versions were unified in a measure called $q$-discord~\cite{costa13}, which is defined as
\be
D_q(\rho)=\min\limits_{\Pi_B}\Big(H_q(\Pi_B[\rho])-H_q(\rho)\Big),
\label{Dq}
\ee
where $H_q(\rho)=\frac{1-\text{Tr}\,\rho^q}{q-1}$ $(q>0\in\mathbb{R})$ is the Tsallis $q$-entropy and $\Pi_B[\rho]=\sum_b\Pi_b\rho\Pi_b$. The entropic (geometric) discord is retrieved from the $q$-discord by setting $q=1$ $(q=2)$. Interestingly, the $q$-discord has been shown to derive solely from the Bayes rule deviation and to keep a link with the couple work-information within the framework of generalized thermodynamic theories~\cite{costa13}. Recently, a conceptual connection has been established between the 1-discord (also called {\em thermal discord}) and elements of reality~\cite{bill15}.

\subsection{Entanglement}

Entanglement reflects the nonseparability of a quantum state, i.e., the fact that a nonseparable state cannot be generated via local operations and classical communication. Among several well-known measures of two-qubit entanglement~\cite{bennett96,wootters98,vidal02} we pick {\em concurrence}~\cite{wootters98},
\be
E(\rho)=\max \left\{ 0,\sqrt{\lambda_1} - \sqrt{\lambda_2} - \sqrt{\lambda_3} - \sqrt{\lambda_4}\right\},
\ee
where $\lambda_i$ are eigenvalues in descending order of the matrix $\rho\tilde\rho$, which is obtained after the computation of the spin-flipped counterpart $\tilde\rho\equiv\sigma_y\otimes\sigma_y\rho^\ast\sigma_y\otimes\sigma_y$ of $\rho$.

\subsection{Steering}

In 2007, Wiseman {\em et al} formalized the concept of {\em steering}~\cite{doherty07}, a term introduced by Schr\"odinger in a reply~\cite{schrodinger35,schrodinger36} to EPR. Steering aims to signalize the presence of quantum correlations that allow Alice to steer Bob's state through her choice of incompatible bases. Later on, a general theory of {\em experimental} steering criteria was developed~\cite{reid09} which revealed its practical appealing, as only a finite number $n$ of measurements are required. Here we stick to the three-measurement criterion, since it has recently been demonstrated that steering based on two measurements actually constitutes a nonlocality measure~\cite{costa16}. We say that a two-qubit system described by a quantum state $\rho$ {\em is non-steerable} in the three-measurement scenario if and only if
\be\label{Iln}
F_3 (\rho,\mu) =\frac{1}{\sqrt{3}} \Big|\sum_{i=1}^3\langle A_i\otimes B_i\rangle\Big| \leqslant 1,
\ee
where $A_i = \hat{u}_i\cdot\vec{\sigma}$,  $B_i = \hat{v}_i\cdot\vec{\sigma}$, $\vec{\sigma}=(\sigma_1,\sigma_2,\sigma_3)$ is a vector composed of the Pauli matrices, $\hat{u}_i\in\mathbb{R}^3$ are unit vectors, $\hat{v}_i\in\mathbb{R}^3$ are orthonormal vectors, $\mu=\{\hat{u}_1,\cdots,\hat{u}_n,\hat{v}_1,\cdots,\hat{v}_n\}$ is the set of measurement directions, $\langle A_i\otimes B_i\rangle=\text{Tr}(\rho A_i\otimes B_i)$, and $\rho\in\mathcal{H}_A\otimes\mathcal{H}_B$ is some bipartite quantum state. A suitable measure of steering in the three-measurement scenario is given by
\be\label{Sgeneral}
S(\rho):=\max\left\{0,\frac{F_3(\rho)-1}{F_3^{\text{\tiny max}}-1}\right\},
\ee 
where $F_3(\rho)=\max_{\mu}F_3(\rho,\mu)$ and $F_3^{\text{\tiny max}}=\max_{\rho}F_n(\rho)$. The inner maximization is taken over all measurement settings $\mu$, while the outer one selects maximal values that are greater than 1. As shown in Ref.~\cite{costa16}, this measure has been defined so as to ensure that $S\in [0,1]$.

\subsection{Bell nonlocality}

Bell nonlocality, the property of a quantum state whose correlations cannot be explained in terms of any local hidden variable model (LHVM), is generally detected by violations of Bell-like inequalities. For two-qubit systems, the Clauser-Horne-Shimony-Holt (CHSH) inequality~\cite{chsh} is of particular interest because of its simplicity. It is written as
\be\label{bell}
B(\rho,\mu):=|\text{Tr}(\rho\,B_{\text{\tiny CHSH}})| \leq 2,
\ee
where $B_{\text{\tiny CHSH}}$, a Bell operator acting on $\mathcal{H}_A\otimes\mathcal{H}_B$ and involving Alice and Bob's measurement settings $\mu$, is conveniently conceived so as to ensure that the above inequality is satisfied for any $\rho$ describable by a LHVM. Our measure of Bell nonlocality is inspired by the results reported in Refs.~\cite{3h95,hu13,costa16} for two-qubit systems, by which one is able to compute $B(\rho):=\max_{\mu}B(\rho,\mu)$. The optimization selects the measurement setting $\mu$ that maximizes nonlocality. Choosing a convenient normalization, we define the Bell nonlocality quantifier as
\be
N(\rho):=\max\left\{0,\frac{B(\rho)-2}{B_{\max}(\rho)-2}\right\}.
\ee
Because $B(\rho)\leqslant B_{\max}(\rho) = 2\sqrt{2}$ (Cirel'son's bound), one has that $N(\rho)\in[0,1]$.

Using the above measures, and their resulting analytical closed formulas for a given class of states, we present in the next section numerical results pointing to a well defined hierarchy. Also, we provide studies of the implied chronology of deaths and births for two-qubit systems under lossy channel in both the Markovian and non-Markovian regimes.

\section{Results \label{results}}

In this work we focus on real two-qubit X states~\cite{yu07} with maximal marginals. This is a class of  three-parameter states that can be written in terms of the Pauli matrices as $\rho_{\vec{c}} = \frac{1}{4}\left(\mathbbm{1}\otimes \mathbbm{1} + \sum_{i=1}^3 c_i\,\sigma_i\otimes\sigma_i\right)$, where $\mathbbm{1}$ is the identity matrix and $\vec{c}=(c_1,c_2,c_3)$ is a real vector with $-1 \leq c_i \leq 1$. In matrix notation, one has that
\be\label{xstates}
\rho_{\vec{c}}=\frac{1}{4}
\begin{pmatrix}
1+c_3 & 0 & 0 & c_1-c_2 \\
0 & 1-c_3 & c_1+c_2 & 0 \\
0 & c_1+c_2 & 1-c_3 & 0 \\
c_1-c_2 & 0 & 0 & 1+c_3
\end{pmatrix}.
\ee
For this class of states, all the quantum correlation measures have been analytically computed in the literature (see, e.g., Refs.\cite{costa16,costa13} and references therein). The results can be expressed as in Table~\ref{table1}. Using these formulas, we proceeded to an exhaustive statistical analysis involving the computation and comparison of these measures over $10^6$ randomly generated vectors $\vec{c}$ for each diagram. 
\begin{table}[htp]
\caption{\small Analytical results for the quantum correlation measures $q$-discord $(D_q)$, entanglement $(E)$, steering $(S)$, and Bell nonlocality $(N)$ for the two-qubit X state $\rho_{\vec{c}}$ given by Eq.~\eqref{xstates}. Here, $\{\lambda_i\}$ are the eigenvalues of $\rho_{\vec{c}}$, $c_{\max}\equiv\max\{|c_1|,|c_2|,|c_3|\}$, $c_{\min}\equiv\min\{|c_1|,|c_2|,|c_3|\}$, and $c = \sqrt{c_1^2+c_2^2+c_3^2}$. A convenient normalization was introduced in the $q$-discord so as to yield $D_q\in[0,1]$.
}
\centering
\begin{tabular}{l}
\hline\hline
Quantum correlation measures\\
\hline 
$ D_q(\rho_{\vec{c}})=\frac{1}{1-2^{1-q}}\left[\sum\limits_{i=1}^4\lambda_i^q-\frac{(1+c_{\max})^q+(1-c_{\max})^q}{2^{2q-1}} \right]$ \\ \\
$ E(\rho_{\vec{c}})=\max\left\{0, \frac{|c_1 - c_2| - |1 - c_3|}{2}, \frac{|c_1 + c_2| - |1 + c_3|}{2} \right\}$\\ \\
$ S(\rho_{\vec{c}})=\max\left\{0,\frac{c-1}{\sqrt{3}-1}\right\}$\\ \\
$ N(\rho_{\vec{c}})=\max\left\{0, \frac{\sqrt{c^2 - c_{\min}^2}-1}{\sqrt{2}-1}\right\}$\\ 
\hline\hline
\end{tabular}\label{table1}
\end{table}
%

\subsection{Equivalence of all $q$-discords}

\begin{figure}[htb]
\centerline{\includegraphics[scale=0.22]{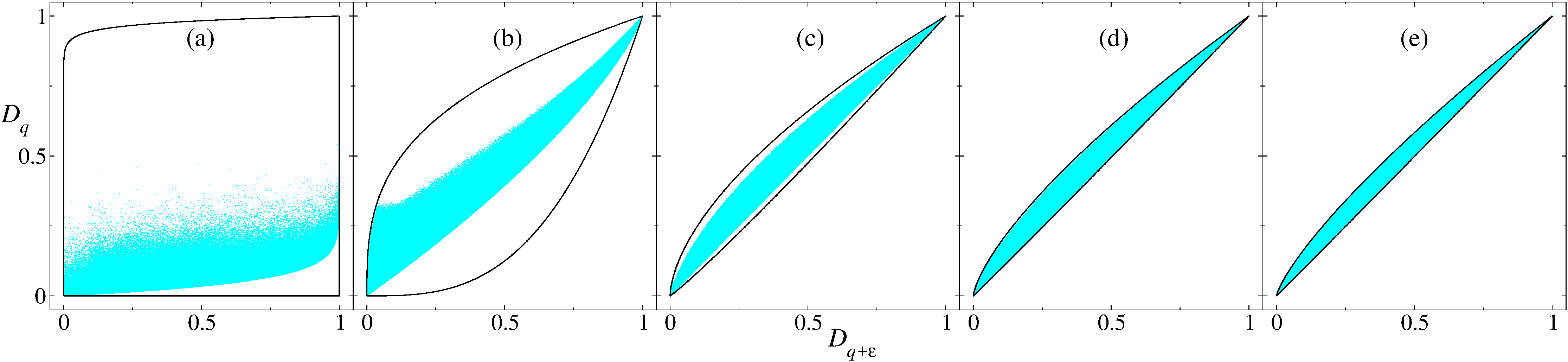}}
\caption{(Color online) Diagrams $D_{q}\times D_{q+\epsilon}$. Each graph contains $10^6$ randomly generated (cyan) points, each one corresponding to a state $\rho_{\vec{c}}$ and a given $\epsilon$ randomly generated within the interval $[0,\epsilon_{\max}]$, with $\epsilon_{\max}=3$. The lower and upper bounds (dashed black lines) are given by the formulas \eqref{bounds}. In these calculations we have employed (a) $q=0.05$, (b) $q=1$, (c) $q=3$, (d) $q=5$, and (e) $q=7$.}
\label{fig2}
\end{figure}

Figure \ref{fig2} shows our numerical results for $D_q\times D_{q+\epsilon}$ diagrams with $0.05\leqslant q\leqslant 7$ and $\epsilon$ a randomly generated real number in $[0,\epsilon_{\max}]$. For these diagrams, we have empirically found the following bounds (represented in Fig.~\ref{fig2} by solid black lines):
\begin{subequations}\label{bounds}
\be 
D_q&=&D_{q+\epsilon}^{\left(\frac{q^s}{q^s+\epsilon_{\max}}\right)}, \qquad \text{with}\,\,s=1\quad \text{(upper bound),} \\ \nonumber \\
D_q&=&D_{q+\epsilon}^{\left(\frac{q^s+\epsilon_{\max}}{q^s}\right)}, \qquad \text{with}\,\,s=4\quad \text{(lower bound).}
\ee
\end{subequations}
We have checked these bounds for a number of values of $q$ and $\epsilon_{\max}$, with $\epsilon\in[0,\epsilon_{\max}]$, having found no violation. As is suggested by the diagrams presented in Fig.~\ref{fig2} and corroborated by the analytical bounds \eqref{bounds}, the lines $(D_{q+\epsilon}>0,0)$ and $(0,D_q>0)$ are excluded from all $D_{q}\times D_{q+\epsilon}$ diagrams. It follows from this analysis and Definition 2, that:

\vskip2mm
\noindent {\bf Result 1}.---{\em All $q$-discords are equivalent for states $\rho_{\vec{c}}$, i.e., $D_q\equiv D_{q+\epsilon}$, $\forall\,(q,\epsilon)>0$.}
\vskip2mm

\noindent This means that $D_q$ and $D_{q'\neq q}$ will always agree whether a given state $\rho_{\vec{c}}$ possesses either null or maximal discord. As a consequence, we do not need to investigate the hierarchy respected by each $q$-discord in relation to the other correlation measures; a single value, say $q=1$, suffices.

\subsection{Hierarchy of quantum resources}

\begin{figure}[htb]
\centerline{\includegraphics[scale=0.23]{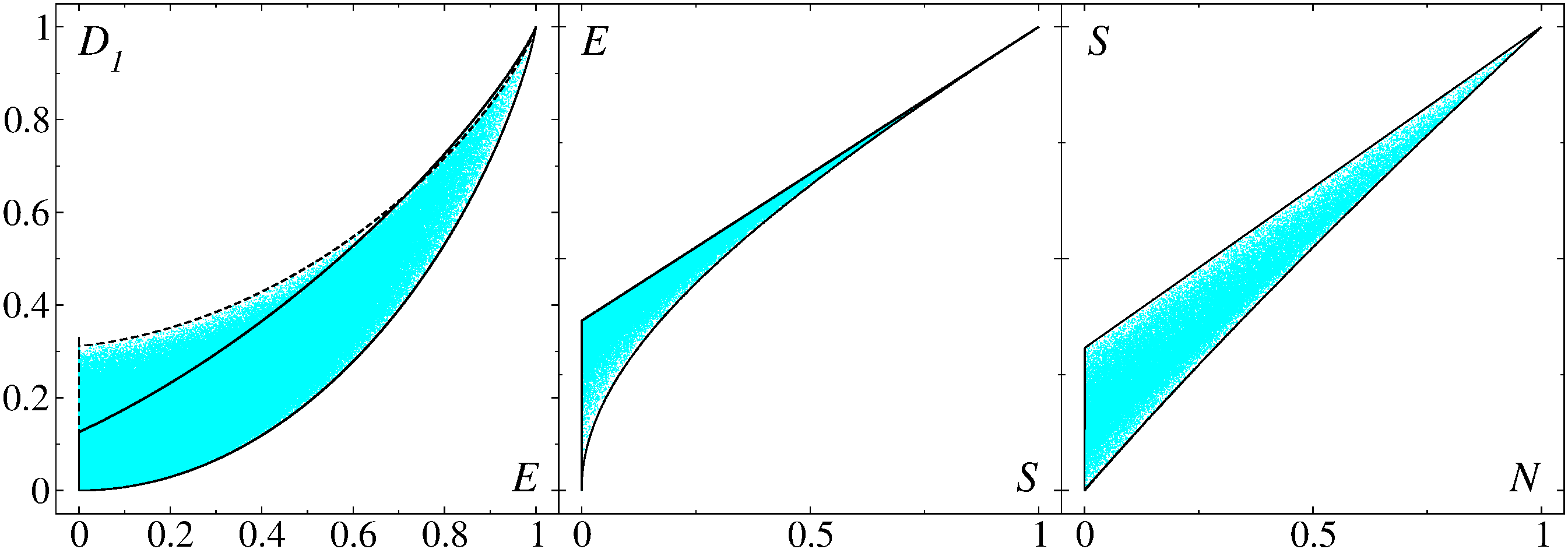}}
\caption{(Color online) Diagrams $D_1 \times E$ (first graph), $E \times S$ (middle graph), and $S \times N$ (third graph). Each graph contains $10^6$ randomly generated (cyan) points, each one corresponding to a state $\rho_{\vec{c}}$. The lower bounds (solid black lines) are given by formulas \eqref{boundsh}. See text for details about the upper bounds.}
\label{fig3}
\end{figure}

Our numerical results for the hierarchy of resources are shown in Fig.~\ref{fig3}. The lower bounds shown in the diagrams correspond to the parametric plot $Q_2\times Q_1$ for the state $\vec{c}=(u,u,-1)$, with $u\in[0,1]$. They can be written as
\begin{subequations}\label{boundsh}
\be 
D_1(E)&=&\frac{2E\arctan{E}+\ln{(1-E^2)}}{\ln{4}}, \\
E(S)&=&\sqrt{\frac{\left[S(\sqrt{3}-1)+1\right]^2-1}{2}}, \\
S(N)&=&\frac{\sqrt{1+2N^2}-1}{\sqrt{3}-1}.
\ee 
\end{subequations}
The upper bounds cannot be exhibited as analytical functions, as they contain the line $(0,Q_2)$, but they can be numerically computed and presented in a parametric plot. While for the diagrams $E\times S$ and $S\times N$ the upper bound is furnished by the Werner state $\vec{c}=-u(1,1,1)$, with $u\in [0,1]$, in $D_1\times E$ the same Werner state competes with the state $\vec{c}=(u,-u,2u-1)$, with $u\in[0,1]$ (represented by a dashed black line).

\noindent These observations allow us to state the following result. 

\vskip2mm
\noindent {\bf Result 2}.---{\em The hierarchy of quantum resources for states $\rho_{\vec{c}}$ is given by $N \triangleright S \triangleright E \triangleright D_1\equiv D_q.$}
\vskip2mm

\noindent The equivalence $D_1\equiv D_q$ comes by Result 1. It is worth noticing that these results are in full agreement with analytical results involving entanglement, steering, and Bell nonlocality reported in Ref.~\cite{costa16}. Also, it is clear that no anomaly appears in the present scenario.

A remark is now opportune. Due to certain degree of arbitrariness involved in the definition of the measures in question, it is not possible to ascribe unambiguous meaning to their absolute values. Then, the shape of the shaded regions in Fig.~\ref{fig3} and their bounds inevitably depend on the chosen definitions. There is, however, one aspect that is invariant, namely, the fact that there should not be any two-qubit X state that is entangled and nondiscordant, steerable and separable, Bell local and non-steerable. The hierarchy expressed by Result 2 reveals quantum resources that are strictly necessary for the existence of others.

\subsection{Chronology of ``deaths'' and ``births'' in noisy channels \label{noise}}

An immediate consequence of the above results is that the quantum resources under concern will accordingly exhibit a hierarchy of {\em robustness} under arbitrary lossy channels. This is so because if $Q_1\triangleright Q_2$ holds, then $Q_2$ cannot vanish before $Q_1$. It follows that $q$-discord is the most resistant resource, successively followed by entanglement, steering, and Bell nonlocality. In other words, it is Bell nonlocality that is expected to be suppressed in first place under noisy channels. Due to the discontinuous character inherent to the maximization procedures in our measures, it is then natural to expect the occurrence of {\em sudden deaths} and {\em sudden births} under lossy dynamics, which must occur according to a chronological ordering obeying the hierarchy $N \triangleright S \triangleright E \triangleright D_q$. In what follows, we verify this point in a concrete scenario.

Consider a situation in which a state $\rho_{\vec{c}}$ is subjected to noisy channels. To describe such a lossy dynamics, we employ the operator-sum representation formalism (Kraus representation)~\cite{chuang}, which implements general effects of independent noisy channels through a mapping $\mathcal{E}$ defined as
\be\label{kraus}
\mathcal{E} (\rho) \equiv \sum_{ij} (K_i\otimes K_j)\, \rho\, (K_i\otimes K_j)^\dagger,
\ee 
where $\sum_i K_i^\dagger K_i = \mathbbm{1}$. $\{K_i\}$ is a set of Kraus operators defining a given trace-preserving quantum operation. Among the variety of known quantum channels, there is a subset that maps X states onto X states. These channels are built with the Kraus sets listed in Table~\ref{table2}. It is just an exercise to show that for those channels it holds that $\mathcal{E}(\rho_{\vec{c}})=\rho_{\vec{c}\,'}$, where $\vec{c}\,'$ is obtained according to the prescriptions presented in Table~\ref{table3}. For simplicity, we restricted ourselves to the case in which both qubits are subjected to the same quantum channel. 
\begingroup
\begin{table}[htb]
\caption{\small Kraus operators for the quantum channels: bit flip (BF), bit-phase flip (BPF), phase flip (PF), depolarizing (DP), and generalized amplitude damping (GAD), where $p$ and $\gamma$ are noise probabilities.}
\centering
\begin{tabular}{cllll}
\hline\hline
Channel & &  Kraus operators\\
\hline\hline
BF  & \hspace{0.75cm} & $K_0=\sqrt{1-p/2}\,\mathbbm{1}$, & \hspace{0.3cm} & $K_1 = \sqrt{p/2}\,\sigma_1$. \\
BPF  &  & $K_0=\sqrt{1-p/2}\,\mathbbm{1}$, &  & $K_1 = \sqrt{p/2}\,\sigma_2$. \\ 
PF &  & $K_0=\sqrt{1-p/2}\,\mathbbm{1}$, &  & $K_1 = \sqrt{p/2}\,\sigma_3$. \\ 
DP & & $K_0=\sqrt{1-3p/4}\,\mathbbm{1}$, & & $K_{1,2,3}=\sqrt{p/4}\,\sigma_{1,2,3}$. \\
GAD &  & $K_0=\frac{1}{\sqrt{2}}\left(\begin{smallmatrix} 1 & 0 \\ 0 & \sqrt{1-\gamma}\end{smallmatrix}\right)$, & & $K_2=\frac{1}{\sqrt{2}}\left(\begin{smallmatrix} \sqrt{1-\gamma} & 0 \\ 0 & 1\end{smallmatrix}\right)$,  \\
& & $K_1=\frac{1}{\sqrt{2}}\left(\begin{smallmatrix} 0 & \sqrt{\gamma} \\ 0 & 0\end{smallmatrix}\right)$, & & $K_3=\frac{1}{\sqrt{2}} \left(\begin{smallmatrix} 0 & 0 \\ \sqrt{\gamma} & 0\end{smallmatrix}\right)$.\\
\hline\hline
\end{tabular}\label{table2}
\end{table}
\endgroup
\begin{table}[htb]
\caption{\small Transformation rules $\vec{c} \stackrel{\text{\tiny $\mathcal{E}$}}{\mapsto} \vec{c}\,'$ emerging from the mapping $\mathcal{E}(\rho_{\vec{c}})=\rho_{\vec{c}\,'}$ associated with the quantum channels: bit flip (BF), bit-phase flip (BPF), phase flip (PF),  depolarizing (DP), and generalized amplitude damping (GAD). Here we consider that both qubits are subjected to the same quantum channel.}
\centering
\begin{tabular}{ccccccc}
\hline \hline
Channel & \hspace{0.75cm} & $c_1'$ & \hspace{0.65cm} & $c_2'$ & \hspace{0.65cm} & $c_3'$ \\
\hline
BF && $c_1$ && $c_2(1-p)^2$ && $c_3(1-p)^2$ \\
BPF && $c_1(1-p)^2$ &&  $c_2$ && $c_3(1-p)^2$ \\
PF && $c_1(1-p)^2$ &&  $c_2(1-p)^2$ && $c_3$ \\
DP && $c_1(1-p)^2$  &&  $c_2(1-p)^2$ && $c_3(1-p)^2$ \\
GAD && $c_1(1-\gamma)$ &&  $c_2(1-\gamma)$ && $c_3(1-\gamma)^2$ \\
\hline\hline
\end{tabular} \label{table3}
\end{table}

The relevance of such maps to the present work relies on the fact that the shaded area in a diagram $Q_2\times Q_1$ constitutes both the domain and the image space for theses maps, so that the resource hierarchy will be respected for all times. This remark, along with the well-known asymptotic behaviour of discord, gives support to the following statement.

\vskip2mm
\noindent {\bf Result 3}.---{\em Let $t_{Q}$ and $\tau_Q$ be the shortest instants of time at which the resource $Q$ vanishes and resurges, respectively, in a lossy dynamics. For states $\rho_{\vec{c}}$, a death chronology and a birth chronology hold such that $t_{D_q}\geqslant t_E \geqslant t_{S} \geqslant t_{N}$, with $t_{D_q}=\infty$, and $\tau_{N}\geqslant \tau_{S}\geqslant \tau_E$.}
\vskip2mm

Next, we provide some illustrations of a strict form of these chronologies by introducing time dependence in the parameters $p$ and $\gamma$.

\subsubsection{Markovian channels}

Here we take the usual parametrization 
\be\label{pgM}
p(t)=\gamma(t)=1-e^{-t/2},
\ee
where $t=\Gamma \tau$ is a dimensionless time scale that makes reference to the physical time $\tau$ in units of a given relaxation time $\Gamma^{-1}$. Without any generality loss, hereafter we will assume that $\Gamma=1$ [a.u.], so that $t$ is equivalent to the physical time $\tau$. For any of the channels given in Table~\ref{table2}, it can be analytically checked from the results presented in Table \ref{table1} that at the equilibrium $(t\to\infty)$ no resource survives, i.e., $\lim_{t\to\infty} Q(\mathcal{E}_t(\rho_{\vec{c}})) = \lim_{t\to\infty} Q(\rho_{\vec{c}\,'(t)})\to 0$. Furthermore, from the results presented in Table~\ref{table3} and the asymptotic behaviour of $p(t)$ and $\gamma(t)$, we see that no recurrence can take place for any resource, which signalizes that the model~\eqref{pgM} suitably implements the physics of Markovian channels. Most importantly, from a thorough analysis of the analytical results of Tables \ref{table1} and \ref{table3} and extensive numerical simulations, we found the following result.

\vskip2mm
\noindent {\bf Result 4}.---{\em Excluding a set of measure zero, which is defined by the pure states 
\be
\vec{c}=\{(-1,-1,-1),(-1,1,1),(1,-1,1),(1,1,-1)\},
\ee
all of which corresponding to the extremal point $(1,1)$ in the diagrams $Q_2\times Q_1$ and for which the death of all resources occur only asymptotically ($t=\infty$) for channels DP and GAD, all states $\rho_{\vec{c}}$ lead to sudden death of entanglement, steering, and nonlocality for all the channels in Table \ref{table2} with the Markovian model \eqref{pgM}.}
\vskip2mm

\begin{figure}[htb]
\centerline{\includegraphics[scale=0.23]{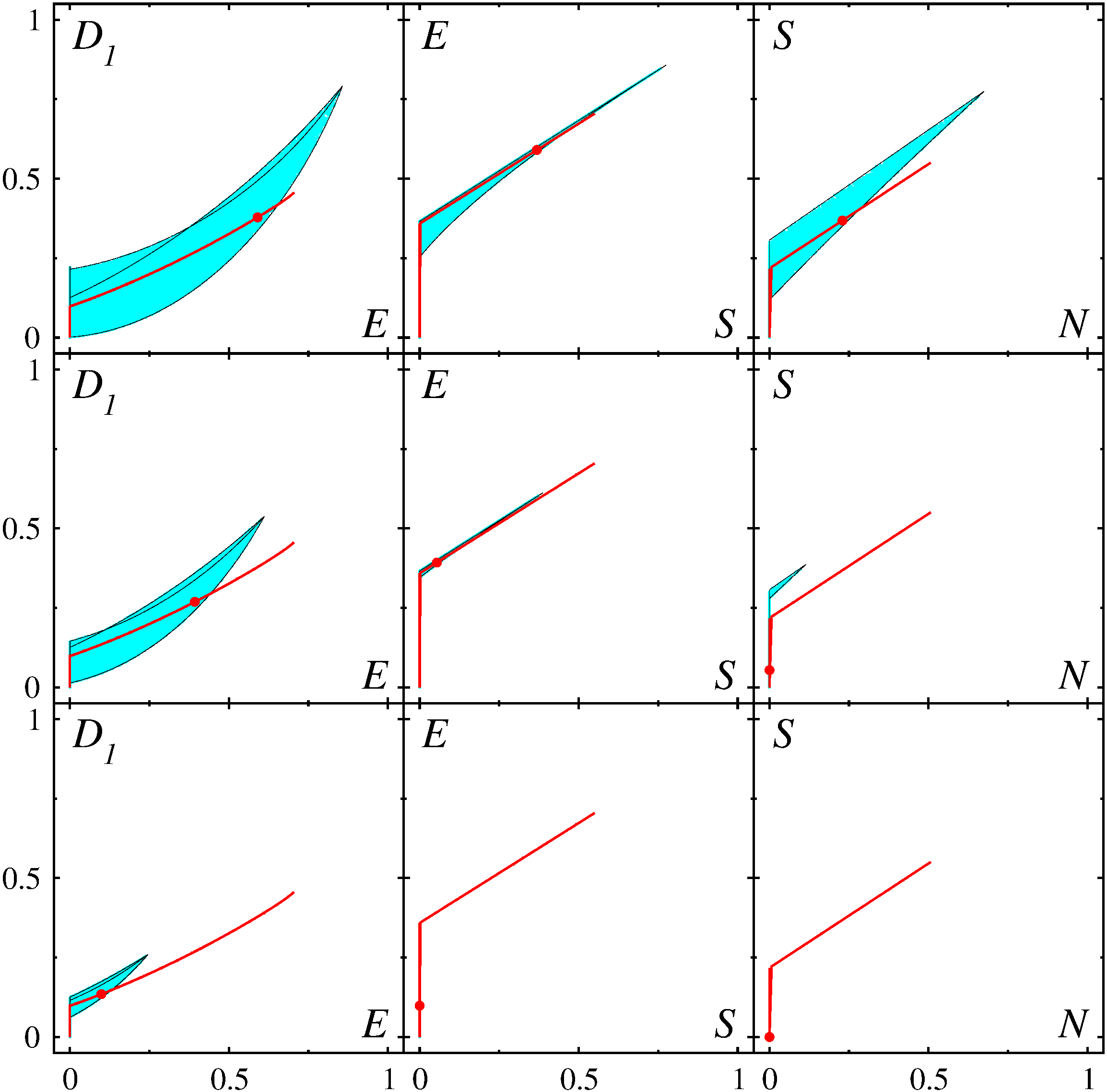}}
\caption{(Color online) Snapshots with $t_1=0.1$ (upper diagrams), $t_2=0.3$ (middle diagrams), and $t_3=0.7$ (lower diagrams) for the time evolution of $55\times 10^4$ states $\rho_{\vec{c}}$ (cyan points) under a Markovian DP channel. At the instant $t_0=0$ the scenario is as in Fig.~\ref{fig3}. The tick red line is the trajectory referring to the state $\vec{c}=(-0.71, -0.75, -0.95)$, whose behaviour is typical, and the red bullet is the position of this state at the instants $t_{1,2,3}$. Notice the sudden death dynamics for all resources except discord.}
\label{fig4}
\end{figure}

\noindent This means that except for a set of measure zero, all states $\rho_{\vec{c}}$ will display abrupt suppression of entanglement, steering, and nonlocality at certain times obeying the chronology stated in Result 3, and asymptotic suppression of discord. This scenario is illustrated in Fig.~\ref{fig4} for a Markovian channel.

\begin{figure}[h!]
\centerline{\includegraphics[scale=0.6]{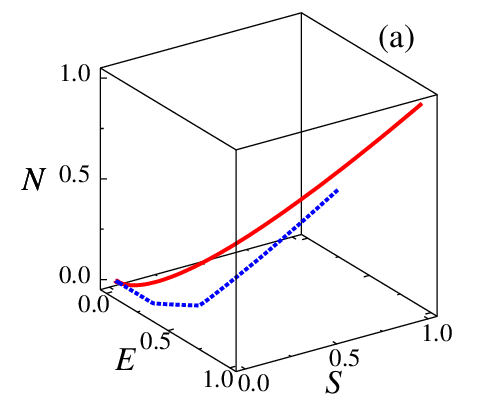}\hskip0.2cm\includegraphics[scale=0.6]{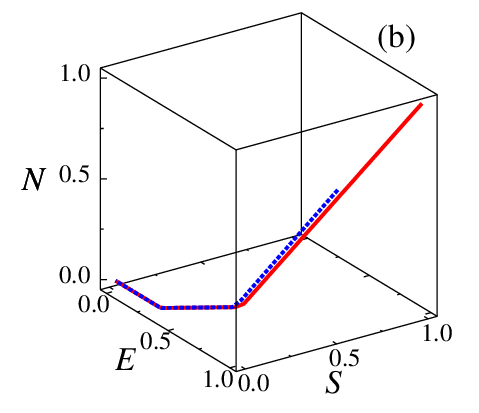}\hskip0.2cm\includegraphics[scale=0.6]{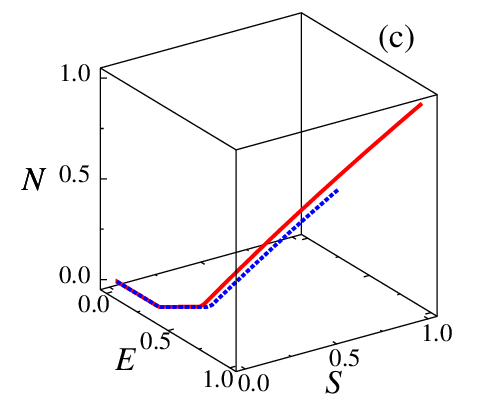}}
\caption{(Color online) Three-dimensional diagram $N\times S\times E$ showing the time evolution of the states $\vec{c}=(-1,-1,-1)$ (solid red line) and $\vec{c}=(0.84,0.91,-0.84)$ (dashed blue line) under channels (a) PF, (b) DP, and (c) GAD. Notice that for the pure state $\vec{c}=(-1,-1,-1)$ under a PF channel the disappearance of resources occur asymptotically (see Result 4).}
\label{fig5}
\end{figure}

Figure \ref{fig5} gives an illustration of two trajectories in three-dimensional diagrams $N\times S\times E$. These trajectories represent the time evolution of two states $\rho_{\vec{c}}$ under three distinct channels, namely, PF, DP, and GAP. Although only two states are shown, these examples are representative of the only two types of behaviour we can have for Markovian trajectories in this context, namely, a trajectory either moves asymptotically to the attractor, as in Fig. \ref{fig5}(a), or displays successive abrupt changes obeying the resource hierarchy, which is the typical behavior.

An illustration of the sudden death chronology can be given for initial states with $\vec{c}=(u,u,v)$, $\tfrac{v-1}{2}\leqslant u\leqslant \tfrac{1-v}{2}$, and $-1\leqslant v\leqslant 1$, under a phase flip (PF) channel. In this case, we can analytically derive the following sudden-death times:
\begin{subequations}
\be
t_E&=&\ln\left(\frac{2|u|}{1+v}\right),\\
t_{S}&=&\frac{1}{2}\ln\left(\frac{2u^2}{1-v^2}\right)=\frac{t_E}{2}-\ln{\sqrt{\frac{(1-v)}{|u|}}},\\
t_N &=&\frac{1}{2}\max\left[\ln\left(\frac{u^2}{1-v^2}\right),\,\ln\left(2u^2\right)\right]=\max\left[t_S-\ln{\sqrt{2}}\,,\,t_S+\ln{\sqrt{1-v^2}}\right].
\ee
\end{subequations}
By the domain of $u$, we see that $\frac{1-v}{|u|}\geqslant 2$, which leads to $t_S\leqslant \frac{t_E}{2} -\ln{\sqrt{2}}$. On the other hand, by the domain of $v$, we see that $\ln\sqrt{1-v^2}$ is strictly non-positive, so that $t_N=t_S-\min[\ln{\sqrt{2}},|\ln{\sqrt{1-v^2}}|]$. From these results it follows the strict chronology $t_N< t_S < t_E$, which is in accordance with Result 3.

\subsubsection{Non-Markovian channels \label{non-markov}}

\begin{figure}[h!]
\centerline{\includegraphics[scale=0.33]{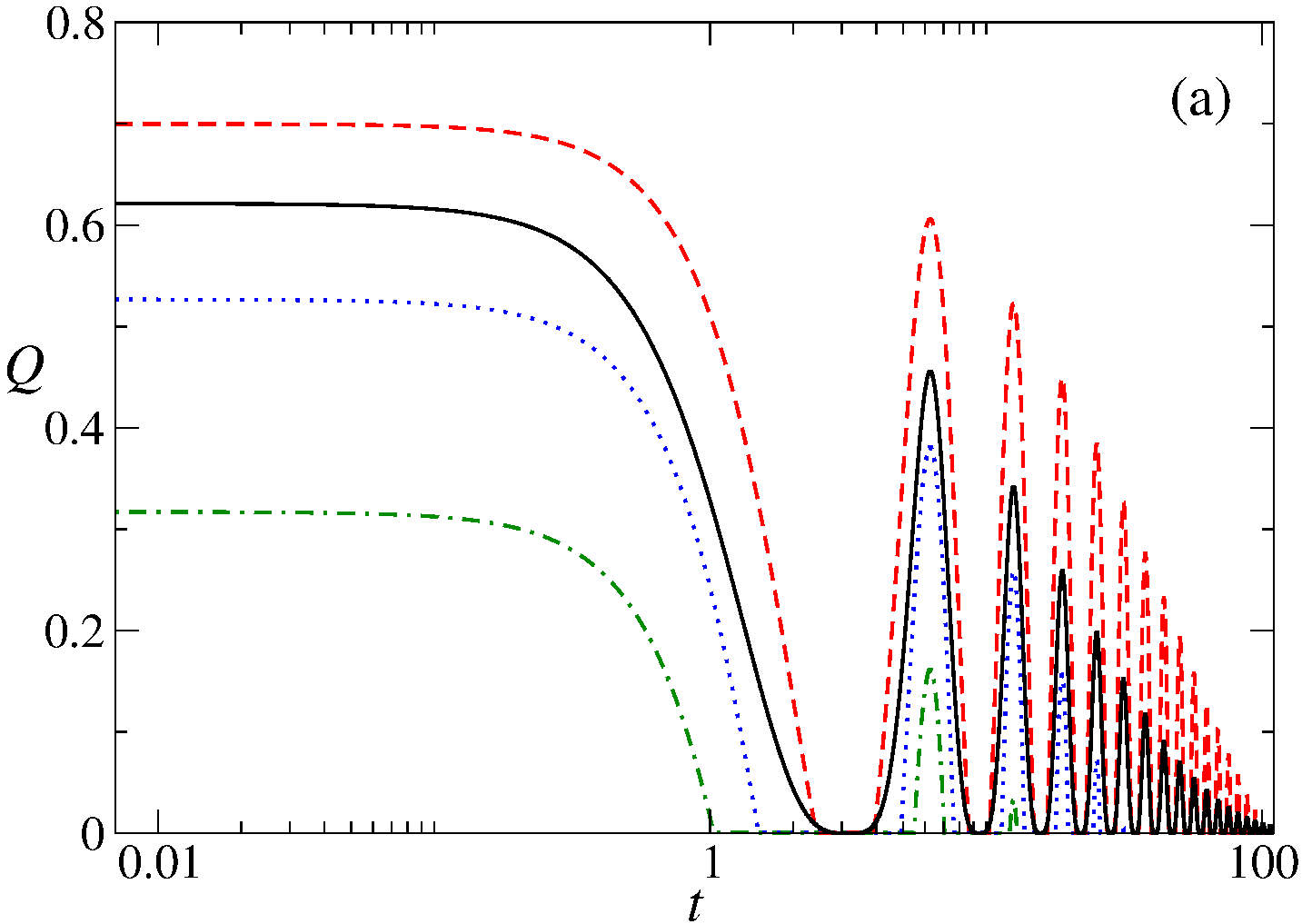}\hskip4mm
\includegraphics[scale=0.33]{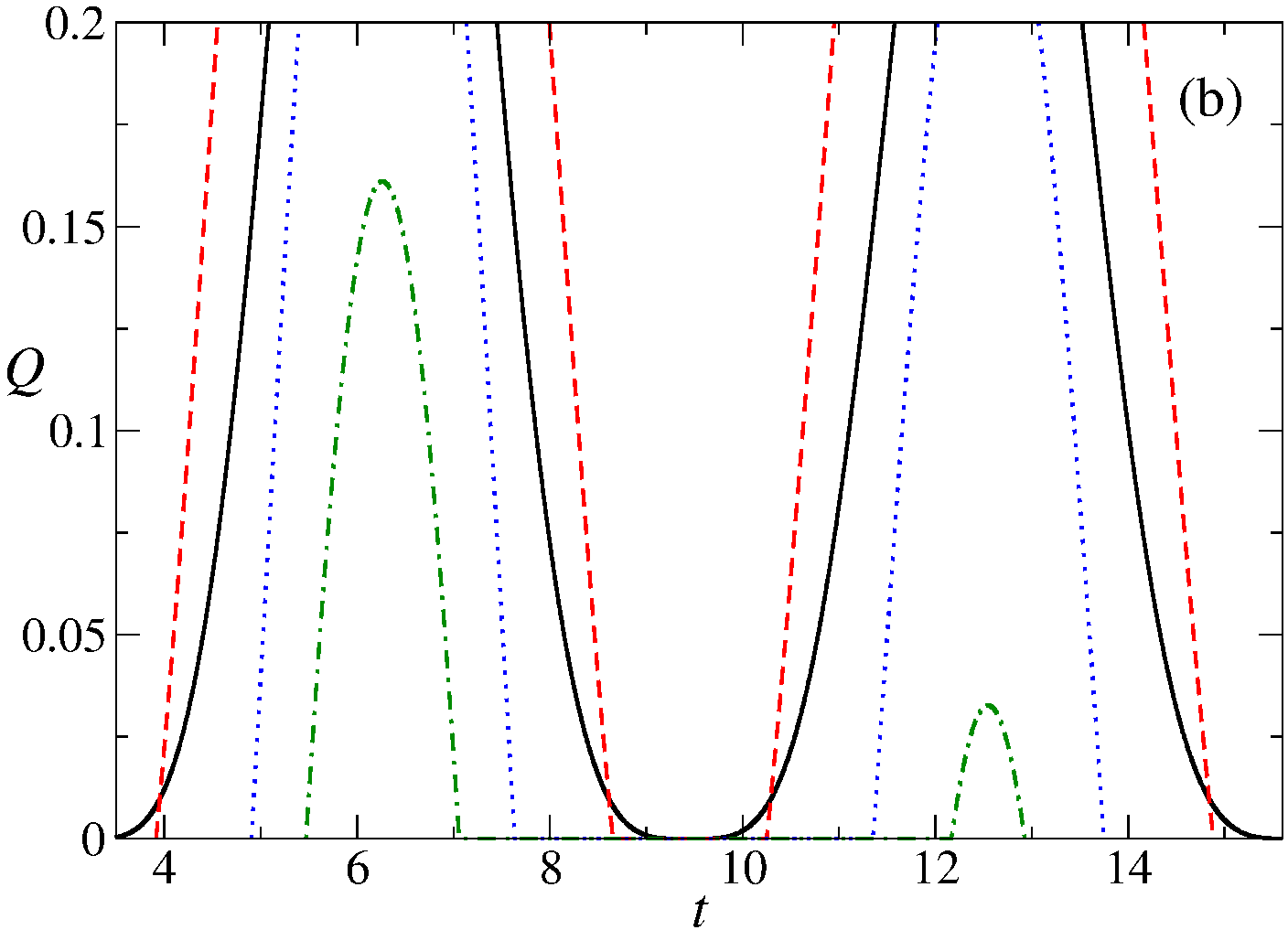}}
\caption{(a) Quantum resources $Q(\rho_{\vec c})$ as a function of the dimensionless time, where $Q=D_1$ (solid black line), $Q=E$ (dashed red line), $Q=S$ (blue dotted line), and $Q=N$ (green dotted-dashed line). These results refer to the dynamics of the Werner state $\vec{c}=(-0.8,-0.8,-0.8)$ in a PF channel with decay rate $\lambda = 0.01$ [a.u.]. Graph (b) is a magnification of a convenient time domain by which we can observe the chronology of deaths and births.}
\label{fig6}
\end{figure}

It is possible to implement non-Markovian dynamics in the formalism by taking the same Kraus operators as before but introducing the following noise probabilities~\cite{bellomo07}:
\be\label{pgNM}
p(t) = \gamma (t) = 1 - e^{-\lambda t}\left[\cos \left(\frac{t}{2}\right) + \lambda \sin \left(\frac{t}{2}\right)\right],
\ee
where $\lambda$ is a decay rate. With this and the analytical results presented in Tables \ref{table1} and \ref{table3}, we can illustrate the birth chronology pointed out in Result 3. In Fig. \ref{fig6} we show the time evolution of a Werner state under a non-Markovian PF channel. The presence of recurrences in the quantum correlation measures $Q$ are signatures of the residual reversibility characteristic of non-Markovian channels. Most interesting, we can clearly see in the magnification provided by Fig. \ref{fig6}(b) the occurrence of both sudden deaths and sudden births of entanglement, steering, and nonlocality, always respecting the chronology demanded by Result 3. Quantum discord, by its turn, disappears and resurges smoothly. Although not shown in the figure, for sufficiently longer times one verifies that $Q(\rho_{\vec{c}})\to 0$.

\section{Summary \label{conclusion}}

In this work, by use of analytical results and convenient statistical tools, such as the resource diagrams, we demonstrated that quantum resources such as $q$-discord, entanglement, steering, and Bell nonlocality obey a hierarchy for the class of real two-qubit X states. This means that for these states, the presence of Bell nonlocality guarantees the presence of steering, and the presence of steering guarantees the presence of entanglement, and entanglement guarantees $q$-discord. In addition, we have shown that all $q$-discord are equivalent to the thermal discord $D_1$.

As a direct consequence of the observed hierarchy, we found a corresponding hierarchy of robustness under generic noisy channels. This implies a necessary chronology of (sudden) deaths and (sudden) births for the resources in question. Discord has shown to be the most robust resource, completely disappearing only asymptotically. Furthermore, we have found that the phenomenon of sudden death of entanglement, steering, and Bell nonlocality actually is {\em the rule rather than the exception} for this class of states. In fact, only a set of states of measure zero does not present sudden death under Markovian channels. Finally, we have also investigated non-Markovian noisy channels and verified the expected chronology of sudden births.

Our findings give strong support to the hierarchy predicted by Refs. \cite{doherty07,doherty07_2,costa16} for entanglement, steering, and nonlocality in scenarios involving few measurements per site. Moreover, we have shown how to locate $q$-discord over this hierarchy. The evidences here provided refer to a particular scenario, so the question naturally arises whether the observed hierarchy of quantum resources will be respected in more general contexts involving several measurements per site, multipartite systems, and Hilbert spaces of higher dimension. This is an open question that delineates an interesting research program.

\section*{Acknowledgments}

This work was supported by CAPES, CNPq, and the National Institute for Science and Technology of Quantum Information (INCT-IQ).

\section*{References}


\begin{thebibliography}{99}

\expandafter\ifx\csname natexlab\endcsname\relax\def\natexlab#1{#1}\fi
\expandafter\ifx\csname bibnamefont\endcsname\relax
  \def\bibnamefont#1{#1}\fi
\expandafter\ifx\csname bibfnamefont\endcsname\relax
  \def\bibfnamefont#1{#1}\fi
\expandafter\ifx\csname citenamefont\endcsname\relax
  \def\citenamefont#1{#1}\fi
\expandafter\ifx\csname url\endcsname\relax
  \def\url#1{\texttt{#1}}\fi
\expandafter\ifx\csname urlprefix\endcsname\relax\def\urlprefix{URL }\fi
\providecommand{\bibinfo}[2]{#2}
\providecommand{\eprint}[2][]{\url{#2}}

\bibitem[{\citenamefont{Einstein et~al}(1935)}]{epr}
\bibinfo{author}{\bibfnamefont{A.} \bibnamefont{Einstein}},
\bibinfo{author}{\bibfnamefont{B.} \bibnamefont{Podolsky}}, \bibnamefont{and}
\bibinfo{author}{\bibfnamefont{N.} \bibnamefont{Rosen}}, 
\bibinfo{title}{\bibfnamefont{Can Quantum-Mechanical Description of Physical Reality Be Considered Complete?}},
  \bibinfo{journal}{Phys.~Rev.} \textbf{\bibinfo{volume}{47}} (\bibinfo{year}{1935})
  \bibinfo{pages}{777}.
  
\bibitem[{\citenamefont{Bell}(1964)}]{bell}
\bibinfo{author}{\bibfnamefont{J.~S.} \bibnamefont{Bell}},
\bibinfo{title}{\bibfnamefont{On the Einstein Podolsky Rosen Paradox}},
  \bibinfo{journal}{Phys.} \textbf{\bibinfo{volume}{1}} (\bibinfo{year}{1964})
  \bibinfo{pages}{195}.

\bibitem[{\citenamefont{Popescu}(1994)}]{popescu94}
\bibinfo{author}{\bibfnamefont{S.} \bibnamefont{Popescu}},
\bibinfo{title}{\bibfnamefont{Bell's inequalities versus teleportation: What is nonlocality?}},
  \bibinfo{journal}{Phys.~Rev.~Lett.} \textbf{\bibinfo{volume}{72}} (\bibinfo{year}{1994}) \bibinfo{pages}{797}.

\bibitem[{\citenamefont{Zurek}(2003)}]{zurek03}
\bibinfo{author}{\bibfnamefont{W.~H.} \bibnamefont{Zurek}},
\bibinfo{title}{\bibfnamefont{Decoherence, einselection, and the quantum origins of the classical}},
  \bibinfo{journal}{Rev.~Mod.~Phys.} \textbf{\bibinfo{volume}{75}} (\bibinfo{year}{2003}) \bibinfo{pages}{715}.
  
\bibitem[{\citenamefont{Angelo}(2015)}]{angelo15}
\bibinfo{author}{\bibfnamefont{R.~M.} \bibnamefont{Angelo}} \bibnamefont{and}
\bibinfo{author}{\bibfnamefont{A.~D.} \bibnamefont{Ribeiro}},
\bibinfo{title}{\bibfnamefont{Wave-particle duality: an information-based approach}},
  \bibinfo{journal}{Found.~Phys.} \textbf{\bibinfo{volume}{45}} (\bibinfo{year}{2015}) \bibinfo{pages}{1407}. 

\bibitem[{\citenamefont{Bilobran}(2015)}]{bill15}
\bibinfo{author}{\bibfnamefont{A.~L.~O.} \bibnamefont{Bilobran}} \bibnamefont{and}
\bibinfo{author}{\bibfnamefont{R.~M.} \bibnamefont{Angelo}},
\bibinfo{title}{\bibfnamefont{A measure of physical reality}},
  \bibinfo{journal}{Europhys.~Lett.} \textbf{\bibinfo{volume}{112}}  (\bibinfo{year}{2015}) \bibinfo{pages}{40005}. 


\bibitem[{\citenamefont{Bennett et~al}(2000)}]{vincenzo}
\bibinfo{author}{\bibfnamefont{C.~H.} \bibnamefont{Bennett}}, \bibnamefont{and}
\bibinfo{author}{\bibfnamefont{D.~P.} \bibnamefont{DiVincenzo}}, 
\bibinfo{title}{\bibfnamefont{Quantum information and computation}},
  \bibinfo{journal}{Nature} \textbf{\bibinfo{volume}{404}} (\bibinfo{year}{2000}) \bibinfo{pages}{247}.

\bibitem[{\citenamefont{Nielsen and Chuang}(2000)}]{chuang}
\bibinfo{author}{\bibfnamefont{M.~A.}~\bibnamefont{Nielsen}} \bibnamefont{and}
\bibinfo{author}{\bibfnamefont{I.~L.}~\bibnamefont{Chuang}},
  \emph{\bibinfo{title}{Quantum Computation and Quantum Information}}
  (\bibinfo{publisher}{Cambridge University Press, Cambridge}, \bibinfo{year}{2000}).

\bibitem[{\citenamefont{Galindo et~al}(2002)}]{delgado02}
\bibinfo{author}{\bibfnamefont{A.} \bibnamefont{Galindo}}, \bibnamefont{and}
\bibinfo{author}{\bibfnamefont{M.~A.} \bibnamefont{Mart\'{i}n-Delgado}}, 
\bibinfo{title}{\bibfnamefont{Information and computation: Classical and quantum aspects}},
  \bibinfo{journal}{Rev.~Mod.~Phys.} \textbf{\bibinfo{volume}{74}} (\bibinfo{year}{2002}) \bibinfo{pages}{347}.

\bibitem[{\citenamefont{Buhrman et~al}(2010)}]{wolf10}
\bibinfo{author}{\bibfnamefont{H.} \bibnamefont{Buhrman}},
\bibinfo{author}{\bibfnamefont{R.} \bibnamefont{Cleve}}, 
\bibinfo{author}{\bibfnamefont{S.} \bibnamefont{Massar}}, \bibnamefont{and}
\bibinfo{author}{\bibfnamefont{R.} \bibnamefont{de Wolf}}, 
\bibinfo{title}{\bibfnamefont{Nonlocality and communication complexity}},
  \bibinfo{journal}{Rev.~Mod.~Phys.} \textbf{\bibinfo{volume}{82}} (\bibinfo{year}{2010}) \bibinfo{pages}{665}.

\bibitem[{\citenamefont{Horodecki et~al}(2009)}]{h409}
\bibinfo{author}{\bibfnamefont{R.} \bibnamefont{Horodecki}},
\bibinfo{author}{\bibfnamefont{P.} \bibnamefont{Horodecki}}, \bibnamefont{and}
\bibinfo{author}{\bibfnamefont{K.} \bibnamefont{Horodecki}}, 
\bibinfo{title}{\bibfnamefont{Quantum entanglement}},
  \bibinfo{journal}{Rev.~Mod.~Phys.} \textbf{\bibinfo{volume}{81}} (\bibinfo{year}{2009}) \bibinfo{pages}{865}.

\bibitem[{\citenamefont{Bennett et~al}(1993)}]{wootters93}
\bibinfo{author}{\bibfnamefont{C.~H.} \bibnamefont{Bennett}},
\bibinfo{author}{\bibfnamefont{G.} \bibnamefont{Brassard}},
\bibinfo{author}{\bibfnamefont{C.} \bibnamefont{Crepeau}}, 
\bibinfo{author}{\bibfnamefont{R.} \bibnamefont{Jozsa}}, 
\bibinfo{author}{\bibfnamefont{A.} \bibnamefont{Peres}},  \bibnamefont{and}
\bibinfo{author}{\bibfnamefont{W.~K.} \bibnamefont{Wootters}}, 
\bibinfo{title}{\bibfnamefont{Teleporting an unknown quantum state via dual classical and Einstein-Podolsky-Rosen channels}},
  \bibinfo{journal}{Phys.~Rev.~Lett.} \textbf{\bibinfo{volume}{70}} (\bibinfo{year}{1993}) \bibinfo{pages}{1895}.

\bibitem[{\citenamefont{Takeda et~al}(2013)}]{furusawa13}
\bibinfo{author}{\bibfnamefont{S.} \bibnamefont{Takeda}},
\bibinfo{author}{\bibfnamefont{T.} \bibnamefont{Mizuta}},
\bibinfo{author}{\bibfnamefont{M.} \bibnamefont{Fuwa}},
\bibinfo{author}{\bibfnamefont{P.} \bibnamefont{van Loock}}, \bibnamefont{and}
\bibinfo{author}{\bibfnamefont{A.} \bibnamefont{Furusawa}}, 
\bibinfo{title}{\bibfnamefont{Deterministic quantum teleportation of photonic quantum bits by a hybrid technique}},
  \bibinfo{journal}{Nature} \textbf{\bibinfo{volume}{500}} (\bibinfo{year}{2013}) \bibinfo{pages}{315}.

\bibitem[{\citenamefont{Steffen et~al}(2013)}]{wallraff13}
\bibinfo{author}{\bibfnamefont{L.} \bibnamefont{Steffen}},
\bibinfo{author}{\bibfnamefont{Y.} \bibnamefont{Salathe}},
\bibinfo{author}{\bibfnamefont{M.} \bibnamefont{Oppliger}}, 
\bibinfo{author}{\bibfnamefont{P.} \bibnamefont{Kurpiers}}, 
\bibinfo{author}{\bibfnamefont{M.} \bibnamefont{Baur}}, 
\bibinfo{author}{\bibfnamefont{C.} \bibnamefont{Lang}}, 
\bibinfo{author}{\bibfnamefont{C.} \bibnamefont{Eichler}}, 
\bibinfo{author}{\bibfnamefont{G.} \bibnamefont{Puebla-Hellmann}},  \bibinfo{author}{\bibfnamefont{A.} \bibnamefont{Fedorov}}, \bibnamefont{and}
\bibinfo{author}{\bibfnamefont{A.} \bibnamefont{Wallraff}}, 
\bibinfo{title}{\bibfnamefont{Deterministic quantum teleportation with feed-forward in a solid state system}},
  \bibinfo{journal}{Nature} \textbf{\bibinfo{volume}{500}} (\bibinfo{year}{2013}) \bibinfo{pages}{319}.

\bibitem[{\citenamefont{Ollivier et~al}(2001)}]{zurek01}
\bibinfo{author}{\bibfnamefont{H.} \bibnamefont{Ollivier}}, \bibnamefont{and}
\bibinfo{author}{\bibfnamefont{W.~H.} \bibnamefont{Zurek}},
\bibinfo{title}{\bibfnamefont{Quantum Discord: A Measure of the Quantumness of Correlations}},
  \bibinfo{journal}{Phys.~Rev.~Lett.} \textbf{\bibinfo{volume}{88}} (\bibinfo{year}{2001}) \bibinfo{pages}{017901}.


\bibitem[{\citenamefont{Henderson et~al}(2001)}]{vedral01}
\bibinfo{author}{\bibfnamefont{L.} \bibnamefont{Henderson}}, \bibnamefont{and}
\bibinfo{author}{\bibfnamefont{V.} \bibnamefont{Vedral}}, 
\bibinfo{title}{\bibfnamefont{Classical, quantum and total correlations}},
  \bibinfo{journal}{J.~Phys.~A} \textbf{\bibinfo{volume}{34}} (\bibinfo{year}{2001}) \bibinfo{pages}{6899}.

\bibitem[{\citenamefont{Dakic et~al}(2012)}]{walther12}
\bibinfo{author}{\bibfnamefont{B.} \bibnamefont{Daki\'{c}}},
\bibinfo{author}{\bibfnamefont{Y.~O.} \bibnamefont{Lipp}},
\bibinfo{author}{\bibfnamefont{X.} \bibnamefont{Ma}}, 
\bibinfo{author}{\bibfnamefont{M.} \bibnamefont{Ringbauer}}, 
\bibinfo{author}{\bibfnamefont{S.} \bibnamefont{Kropatschek}}, 
\bibinfo{author}{\bibfnamefont{S.} \bibnamefont{Barz}}, 
\bibinfo{author}{\bibfnamefont{T.} \bibnamefont{Paterek}}, 
\bibinfo{author}{\bibfnamefont{V.} \bibnamefont{Vedral}},
\bibinfo{author}{\bibfnamefont{A.} \bibnamefont{Zeilinger}},
\bibinfo{author}{\bibfnamefont{C.} \bibnamefont{Brukner}}, \bibnamefont{and}
\bibinfo{author}{\bibfnamefont{P.} \bibnamefont{Walther}}, 
\bibinfo{title}{\bibfnamefont{Quantum discord as resource for remote state preparation}},
  \bibinfo{journal}{Nature Phys.} \textbf{\bibinfo{volume}{8}} (\bibinfo{year}{2012}) \bibinfo{pages}{666}.


\bibitem[{\citenamefont{Gu et~al}(2012)}]{lam12}
\bibinfo{author}{\bibfnamefont{M.} \bibnamefont{Gu}},
\bibinfo{author}{\bibfnamefont{H.~M.} \bibnamefont{Chrzanowski}},
\bibinfo{author}{\bibfnamefont{S.~M.} \bibnamefont{Assad}}, 
\bibinfo{author}{\bibfnamefont{T.} \bibnamefont{Symul}}, 
\bibinfo{author}{\bibfnamefont{K.} \bibnamefont{Modi}}, 
\bibinfo{author}{\bibfnamefont{T.~C.} \bibnamefont{Ralph}}, 
\bibinfo{author}{\bibfnamefont{V.} \bibnamefont{Vedral}},  \bibnamefont{and}
\bibinfo{author}{\bibfnamefont{P.~K.} \bibnamefont{Lam}}, 
\bibinfo{title}{\bibfnamefont{Observing the operational significance of discord consumption}},
  \bibinfo{journal}{Nature Phys.} \textbf{\bibinfo{volume}{8}} (\bibinfo{year}{2012}) \bibinfo{pages}{671}.

\bibitem[{\citenamefont{Brodutch et~al}(2013)}]{brodutch13}
\bibinfo{author}{\bibfnamefont{A.} \bibnamefont{Brodutch}},
\bibinfo{title}{\bibfnamefont{Discord and quantum computational resources}},
  \bibinfo{journal}{Phys.~Rev.~A} \textbf{\bibinfo{volume}{88}} (\bibinfo{year}{2013}) \bibinfo{pages}{022307}.

\bibitem[{\citenamefont{Werlang et~al}(2009)}]{boas09}
\bibinfo{author}{\bibfnamefont{T.} \bibnamefont{Werlang}},
\bibinfo{author}{\bibfnamefont{S.} \bibnamefont{Souza}}, 
\bibinfo{author}{\bibfnamefont{F.~F.} \bibnamefont{Fanchini}}, \bibnamefont{and}
\bibinfo{author}{\bibfnamefont{C.~J.} \bibnamefont{Villas Boas}}, 
\bibinfo{title}{\bibfnamefont{Robustness of quantum discord to sudden death}},
  \bibinfo{journal}{Phys.~Rev.~A} \textbf{\bibinfo{volume}{80}} (\bibinfo{year}{2009}) \bibinfo{pages}{024103}.

\bibitem[{\citenamefont{Merali et~al}(2011)}]{merali11}
\bibinfo{author}{\bibfnamefont{Z.} \bibnamefont{Merali}},
\bibinfo{title}{\bibfnamefont{Quantum computing: The power of discord}},
  \bibinfo{journal}{Nature} \textbf{\bibinfo{volume}{474}} (\bibinfo{year}{2011}) \bibinfo{pages}{24}.

\bibitem[{\citenamefont{Celeri et~al}(2011)}]{celeri11}
\bibinfo{author}{\bibfnamefont{L.~C.} \bibnamefont{Celeri}},
\bibinfo{author}{\bibfnamefont{J.} \bibnamefont{Maziero}}, \bibnamefont{and}
\bibinfo{author}{\bibfnamefont{R.~M.} \bibnamefont{Serra}}, 
\bibinfo{title}{\bibfnamefont{Theoretical and experimental aspects of quantum discord and related measures}},
  \bibinfo{journal}{Int.~J.~Quant.~Inf.} \textbf{\bibinfo{volume}{9}} (\bibinfo{year}{2011}) \bibinfo{pages}{1837}.

\bibitem[{\citenamefont{Modi et~al}(2012)}]{vedral12}
\bibinfo{author}{\bibfnamefont{K.} \bibnamefont{Modi}},
\bibinfo{author}{\bibfnamefont{A.} \bibnamefont{Brodutch}},
\bibinfo{author}{\bibfnamefont{H.} \bibnamefont{Cable}}, 
\bibinfo{author}{\bibfnamefont{T.} \bibnamefont{Paterek}},  \bibnamefont{and}
\bibinfo{author}{\bibfnamefont{V.} \bibnamefont{Vedral}}, 
\bibinfo{title}{\bibfnamefont{The classical-quantum boundary for correlations: Discord and related measures}},
  \bibinfo{journal}{Rev.~Mod.~Phys.} \textbf{\bibinfo{volume}{84}} (\bibinfo{year}{2012}) \bibinfo{pages}{1655}.

\bibitem[{\citenamefont{Schrodinger et~al}(1935)}]{schrodinger35}
\bibinfo{author}{\bibfnamefont{E.} \bibnamefont{Schr\"{o}dinger}},
\bibinfo{title}{\bibfnamefont{Discussion of Probability Relations Between Separated Systems}},
  \bibinfo{journal}{Proc.~Cambridge Philos.~Soc.} \textbf{\bibinfo{volume}{31}} (\bibinfo{year}{1935}) \bibinfo{pages}{553}.

\bibitem[{\citenamefont{Schrodinger et~al}(1936)}]{schrodinger36}
\bibinfo{author}{\bibfnamefont{E.} \bibnamefont{Schr\"{o}dinger}},
\bibinfo{title}{\bibfnamefont{Probability relations between separated systems}},
  \bibinfo{journal}{Proc.~Cambridge Philos.~Soc.} \textbf{\bibinfo{volume}{32}} (\bibinfo{year}{1936}) \bibinfo{pages}{446}.

\bibitem[{\citenamefont{Wiseman et~al}(2007)}]{doherty07}
\bibinfo{author}{\bibfnamefont{H.~M.} \bibnamefont{Wiseman}},
\bibinfo{author}{\bibfnamefont{S.~J.} \bibnamefont{Jones}}, \bibnamefont{and}
\bibinfo{author}{\bibfnamefont{A.~C.} \bibnamefont{Doherty}}, 
\bibinfo{title}{\bibfnamefont{Steering, Entanglement, Nonlocality, and the Einstein-Podolsky-Rosen Paradox}},
  \bibinfo{journal}{Phys.~Rev.~Lett.} \textbf{\bibinfo{volume}{98}} (\bibinfo{year}{2007}) \bibinfo{pages}{140402}.

\bibitem[{\citenamefont{Jones et~al}()}]{doherty07_2}
\bibinfo{author}{\bibfnamefont{S.~J.} \bibnamefont{Jones}},
\bibinfo{author}{\bibfnamefont{H.~M.} \bibnamefont{Wiseman}}, \bibnamefont{and}
\bibinfo{author}{\bibfnamefont{A.~C.} \bibnamefont{Doherty}}, 
\bibinfo{title}{\bibfnamefont{Entanglement, Einstein-Podolsky-Rosen correlations, Bell nonlocality, and steering}},
  \bibinfo{journal}{Phys.~Rev.~A} \textbf{\bibinfo{volume}{76}} (\bibinfo{year}{2007}) \bibinfo{pages}{052116}.

\bibitem[{\citenamefont{Cavalcanti et~al}(2009)}]{reid09}
\bibinfo{author}{\bibfnamefont{E.~G.} \bibnamefont{Cavalcanti}},
\bibinfo{author}{\bibfnamefont{S.~J.} \bibnamefont{Jones}},
\bibinfo{author}{\bibfnamefont{H.~M.} \bibnamefont{Wiseman}},  \bibnamefont{and}
\bibinfo{author}{\bibfnamefont{M.~D.} \bibnamefont{Reid}}, 
\bibinfo{title}{\bibfnamefont{Experimental criteria for steering and the Einstein-Podolsky-Rosen paradox}},
  \bibinfo{journal}{Phys.~Rev.~A} \textbf{\bibinfo{volume}{80}} (\bibinfo{year}{2009}) \bibinfo{pages}{032112}.

\bibitem[{\citenamefont{Costa et~al}(2016)}]{costa16}
\bibinfo{author}{\bibfnamefont{A.~C.~S.} \bibnamefont{Costa}}, \bibnamefont{and}
\bibinfo{author}{\bibfnamefont{R.~M.} \bibnamefont{Angelo}}, 
\bibinfo{title}{\bibfnamefont{Quantification of Einstein-Podolski-Rosen steering for two-qubit states}},
  \bibinfo{journal}{Phys. Rev. A} \textbf{\bibinfo{volume}{93}} (\bibinfo{year}{2016}) \bibinfo{pages}{020103(R)}.

\bibitem[{\citenamefont{Law et al}(2014)}]{law14}
\bibinfo{author}{\bibfnamefont{Y. Z.} \bibnamefont{Law}}, 
\bibinfo{author}{\bibfnamefont{L. P.} \bibnamefont{Thinh}},
\bibinfo{author}{\bibfnamefont{J.-D.} \bibnamefont{Bancal}}, \bibnamefont{and}
\bibinfo{author}{\bibfnamefont{V.} \bibnamefont{Scarani}}, 
\bibinfo{author}{\bibfnamefont{Quantum randomness extraction for various levels of characterization of the devices}},
  \bibinfo{journal}{J. Phys. A: Math. Theor.} \textbf{\bibinfo{volume}{47}} (\bibinfo{year}{2014}) \bibinfo{pages}{424028}.

\bibitem[{\citenamefont{Piani et al}(2011)}]{piani15}
\bibinfo{author}{\bibfnamefont{M.} \bibnamefont{Piani}}, \bibnamefont{and}
\bibinfo{author}{\bibfnamefont{J.} \bibnamefont{Watrous}}, 
\bibinfo{author}{\bibfnamefont{Necessary and sufficient quantum information characterization of Einstein-Podolsky-Rosen steering}},
  \bibinfo{journal}{Phys. Rev. Lett. } \textbf{\bibinfo{volume}{114}} (\bibinfo{year}{2015}) \bibinfo{pages}{060404}. 
  
\bibitem[{\citenamefont{Branciard et al}(2011)}]{branciard11}
\bibinfo{author}{\bibfnamefont{C.} \bibnamefont{Branciard}} \bibnamefont{and}
\bibinfo{author}{\bibfnamefont{N.} \bibnamefont{Gisin}}, 
\bibinfo{author}{\bibfnamefont{Quantifying the nonlocality of Greenberger-Horne-Zeilinger quantum correlations by a bounded communication simulation protocol}},
  \bibinfo{journal}{Phys. Rev. Lett. } \textbf{\bibinfo{volume}{107}} (\bibinfo{year}{2011}) \bibinfo{pages}{020401}.  
  
\bibitem[{\citenamefont{Branciard et al}(2012)}]{branciard12}
\bibinfo{author}{\bibfnamefont{C.} \bibnamefont{Branciard}}, 
\bibinfo{author}{\bibfnamefont{E. G.} \bibnamefont{Cavalcanti}}, 
\bibinfo{author}{\bibfnamefont{S. P.} \bibnamefont{Walborn}},  
\bibinfo{author}{\bibfnamefont{V.} \bibnamefont{Scarani}}, \bibnamefont{and}
\bibinfo{author}{\bibfnamefont{H. M.} \bibnamefont{Wiseman}}, 
\bibinfo{author}{\bibfnamefont{One-sided device-independent quantum key distribution: security, feasibility, and the connection with steering}},
  \bibinfo{journal}{Phys. Rev. A} \textbf{\bibinfo{volume}{85}} (\bibinfo{year}{2012}) \bibinfo{pages}{010301(R)}.
  
\bibitem[{\citenamefont{Gallego et al}(2015)}]{gallego15}
\bibinfo{author}{\bibfnamefont{R.} \bibnamefont{Gallego}} \bibnamefont{and}
\bibinfo{author}{\bibfnamefont{L.} \bibnamefont{Aolita}}, 
\bibinfo{author}{\bibfnamefont{Resource theory of steering}},
  \bibinfo{journal}{Phys. Rev. X} \textbf{\bibinfo{volume}{5}} (\bibinfo{year}{2015}) \bibinfo{pages}{041008}. 
  
  \bibitem[{\citenamefont{Saunders et al}(2010)}]{saunders10}
\bibinfo{author}{\bibfnamefont{D. J.} \bibnamefont{Saunders}}, 
\bibinfo{author}{\bibfnamefont{S. J.} \bibnamefont{Jones}}, 
\bibinfo{author}{\bibfnamefont{H. M.} \bibnamefont{Wiseman}}, \bibnamefont{and}
\bibinfo{author}{\bibfnamefont{G. J.} \bibnamefont{Pryde}}, 
\bibinfo{author}{\bibfnamefont{Experimental EPR-steering using Bell-local states}},
  \bibinfo{journal}{Nature Phys.} \textbf{\bibinfo{volume}{6}} (\bibinfo{year}{2010}) \bibinfo{pages}{845}.  
  
\bibitem[{\citenamefont{Handchen et al}(2012)}]{handchen12}
\bibinfo{author}{\bibfnamefont{V.} \bibnamefont{Handchen}}, 
\bibinfo{author}{\bibfnamefont{T.} \bibnamefont{Eberle}}, 
\bibinfo{author}{\bibfnamefont{S.} \bibnamefont{Steinlechner}}, 
\bibinfo{author}{\bibfnamefont{A.} \bibnamefont{Samblowski}}, 
\bibinfo{author}{\bibfnamefont{T.} \bibnamefont{Franz}}, 
\bibinfo{author}{\bibfnamefont{R. F.} \bibnamefont{Werner}}, \bibnamefont{and}
\bibinfo{author}{\bibfnamefont{R.} \bibnamefont{Schnabel}}, 
\bibinfo{author}{\bibfnamefont{Observation of one-way Einstein-Podolsky-Rosen steering}},
  \bibinfo{journal}{Nature Phot.} \textbf{\bibinfo{volume}{6}} (\bibinfo{year}{2012}) \bibinfo{pages}{596}.

\bibitem[{\citenamefont{Wittmann et al}(2012)}]{wittmann12}
\bibinfo{author}{\bibfnamefont{B.} \bibnamefont{Wittmann}}, 
\bibinfo{author}{\bibfnamefont{S.} \bibnamefont{Ramelow}}, 
\bibinfo{author}{\bibfnamefont{F.} \bibnamefont{Steinlechner}},  
\bibinfo{author}{\bibfnamefont{N. K.} \bibnamefont{Langford}}, 
\bibinfo{author}{\bibfnamefont{N.} \bibnamefont{Brunner}},
\bibinfo{author}{\bibfnamefont{H. M.} \bibnamefont{Wiseman}}, 
\bibinfo{author}{\bibfnamefont{R.} \bibnamefont{Ursin}}, \bibnamefont{and}
\bibinfo{author}{\bibfnamefont{A.} \bibnamefont{Zeilinger}}, 
\bibinfo{author}{\bibfnamefont{Loophole-free Einstein-Podolsky-Rosen experiment via quantum steering}},
  \bibinfo{journal}{New J. Phys.} \textbf{\bibinfo{volume}{14}} (\bibinfo{year}{2012}) \bibinfo{pages}{053030}.

\bibitem[{\citenamefont{Smith et al}(2012)}]{smith12}
\bibinfo{author}{\bibfnamefont{D. H.} \bibnamefont{Smith}}, 
\bibinfo{author}{\bibfnamefont{G.} \bibnamefont{Gillett}}, 
\bibinfo{author}{\bibfnamefont{M. P.} \bibnamefont{de Almeida}},  
\bibinfo{author}{\bibfnamefont{C.} \bibnamefont{Branciard}}, 
\bibinfo{author}{\bibfnamefont{A.} \bibnamefont{Fedrizzi}},
\bibinfo{author}{\bibfnamefont{T. J.} \bibnamefont{Weinhold}}, 
\bibinfo{author}{\bibfnamefont{A.} \bibnamefont{Lita}},
\bibinfo{author}{\bibfnamefont{B.} \bibnamefont{Calkins}},
\bibinfo{author}{\bibfnamefont{T.} \bibnamefont{Gerrits}}, 
\bibinfo{author}{\bibfnamefont{H. M.} \bibnamefont{Wiseman}},
\bibinfo{author}{\bibfnamefont{S. W.} \bibnamefont{Nam}}, \bibnamefont{and}
\bibinfo{author}{\bibfnamefont{A. G.} \bibnamefont{White}}, 
\bibinfo{author}{\bibfnamefont{Conclusive quantum steering with superconducting transition-edge sensors}},
  \bibinfo{journal}{Nature Commun.} \textbf{\bibinfo{volume}{3}} (\bibinfo{year}{2012}) \bibinfo{pages}{625}.

\bibitem[{\citenamefont{Bennet et al}(2012)}]{bennet12}
\bibinfo{author}{\bibfnamefont{A. J.} \bibnamefont{Bennet}}, 
\bibinfo{author}{\bibfnamefont{D. A.} \bibnamefont{Evans}}, 
\bibinfo{author}{\bibfnamefont{D. J.} \bibnamefont{Saunders}},  
\bibinfo{author}{\bibfnamefont{C.} \bibnamefont{Branciard}}, 
\bibinfo{author}{\bibfnamefont{E. G.} \bibnamefont{Cavalcanti}}, 
\bibinfo{author}{\bibfnamefont{H. M.} \bibnamefont{Wiseman}}, \bibnamefont{and}
\bibinfo{author}{\bibfnamefont{G. J.} \bibnamefont{Pryde}}, 
\bibinfo{author}{\bibfnamefont{Arbitrarily Loss-Tolerant Einstein-Podolsky-Rosen Steering Allowing a Demonstration over 1 km of Optical Fiber with No Detection Loophole}},
  \bibinfo{journal}{Phys. Rev. X} \textbf{\bibinfo{volume}{2}} (\bibinfo{year}{2012}) \bibinfo{pages}{031003}.
  
\bibitem[{\citenamefont{Brunner et al}(2014)}]{brunner14}
\bibinfo{author}{\bibfnamefont{N.} \bibnamefont{Brunner}},
\bibinfo{author}{\bibfnamefont{D.} \bibnamefont{Cavalcanti}},
\bibinfo{author}{\bibfnamefont{S.} \bibnamefont{Pironio}},
\bibinfo{author}{\bibfnamefont{V.} \bibnamefont{Scarani}}, \bibnamefont{and}
\bibinfo{author}{\bibfnamefont{S.} \bibnamefont{Wehner}}, 
\bibinfo{author}{\bibfnamefont{Bell nonlocality,}}
 \bibinfo{journal}{Rev. Mod. Phys.} \textbf{\bibinfo{volume}{86}} (\bibinfo{year}{2014}) \bibinfo{pages}{419}. 


 \bibitem[{\citenamefont{Clauser et~al}(1969)}]{chsh}
\bibinfo{author}{\bibfnamefont{J.~F.} \bibnamefont{Clauser}},
\bibinfo{author}{\bibfnamefont{M.~A.} \bibnamefont{Horne}}, 
\bibinfo{author}{\bibfnamefont{A.} \bibnamefont{Shimony}}, \bibnamefont{and}
\bibinfo{author}{\bibfnamefont{R.~A.} \bibnamefont{Holt}}, 
\bibinfo{title}{\bibfnamefont{Proposed Experiment to Test Local Hidden-Variable Theories}},
  \bibinfo{journal}{Phys.~Rev.~Lett.} \textbf{\bibinfo{volume}{23}} (\bibinfo{year}{1969}) \bibinfo{pages}{880}.

\bibitem[{\citenamefont{Acin et~al}(2007)}]{scarani07}
\bibinfo{author}{\bibfnamefont{A.} \bibnamefont{Ac\'in}},
\bibinfo{author}{\bibfnamefont{N.} \bibnamefont{Brunner}},
\bibinfo{author}{\bibfnamefont{N.} \bibnamefont{Gisin}},
\bibinfo{author}{\bibfnamefont{S.} \bibnamefont{Massar}}, 
\bibinfo{author}{\bibfnamefont{S.} \bibnamefont{Pironio}},  \bibnamefont{and}
\bibinfo{author}{\bibfnamefont{V.} \bibnamefont{Scarani}}, 
\bibinfo{title}{\bibfnamefont{Device-Independent Security of Quantum Cryptography against Collective Attacks}},
  \bibinfo{journal}{Phys.~Rev.~Lett.} \textbf{\bibinfo{volume}{98}} (\bibinfo{year}{2007}) \bibinfo{pages}{230501}.


\bibitem[{\citenamefont{Acin et~al}(2006)}]{masanes06}
\bibinfo{author}{\bibfnamefont{A.} \bibnamefont{Ac\'in}},
\bibinfo{author}{\bibfnamefont{N.} \bibnamefont{Gisin}}, \bibnamefont{and}
\bibinfo{author}{\bibfnamefont{L.} \bibnamefont{Masanes}}, 
\bibinfo{title}{\bibfnamefont{From Bell's Theorem to Secure Quantum Key Distribution}},
  \bibinfo{journal}{Phys.~Rev.~Lett.} \textbf{\bibinfo{volume}{97}} (\bibinfo{year}{2006}) \bibinfo{pages}{120405}.

\bibitem[{\citenamefont{Barrett et~al}(2005)}]{kent05}
\bibinfo{author}{\bibfnamefont{J.} \bibnamefont{Barrett}},
\bibinfo{author}{\bibfnamefont{L.} \bibnamefont{Hardy}}, \bibnamefont{and}
\bibinfo{author}{\bibfnamefont{A.} \bibnamefont{Kent}}, 
\bibinfo{title}{\bibfnamefont{No Signaling and Quantum Key Distribution}},
  \bibinfo{journal}{Phys.~Rev.~Lett.} \textbf{\bibinfo{volume}{95}} (\bibinfo{year}{2005}) \bibinfo{pages}{010503}.


\bibitem[{\citenamefont{Pironio et~al}(2010)}]{monroe10}
\bibinfo{author}{\bibfnamefont{S.} \bibnamefont{Pironio}},
\bibinfo{author}{\bibfnamefont{A.} \bibnamefont{Ac\'in}}, 
\bibinfo{author}{\bibfnamefont{S.} \bibnamefont{Massar}}, 
\bibinfo{author}{\bibfnamefont{A.} \bibnamefont{Boyer de la Giroday}}, 
\bibinfo{author}{\bibfnamefont{D.~N.} \bibnamefont{Matsukevich}}, 
\bibinfo{author}{\bibfnamefont{P.} \bibnamefont{Maunz}}, 
\bibinfo{author}{\bibfnamefont{S.} \bibnamefont{Olmschenk}}, 
\bibinfo{author}{\bibfnamefont{D.} \bibnamefont{Hayes}}, 
\bibinfo{author}{\bibfnamefont{L.} \bibnamefont{Luo}}, 
\bibinfo{author}{\bibfnamefont{T.~A.} \bibnamefont{Manning}}, \bibnamefont{and}
\bibinfo{author}{\bibfnamefont{C.} \bibnamefont{Monroe}}, 
\bibinfo{title}{\bibfnamefont{Random numbers certified by Bell's theorem}},
  \bibinfo{journal}{Nature} \textbf{\bibinfo{volume}{464}} (\bibinfo{year}{2010}) \bibinfo{pages}{1021}.


\bibitem[{\citenamefont{Brukner et~al}(2004)}]{zeilinger04}
\bibinfo{author}{\bibfnamefont{C.} \bibnamefont{Brukner}},
\bibinfo{author}{\bibfnamefont{M.} \bibnamefont{Zukowski}},
\bibinfo{author}{\bibfnamefont{J.~W.} \bibnamefont{Pan}},  \bibnamefont{and}
\bibinfo{author}{\bibfnamefont{A.} \bibnamefont{Zeilinger}}, 
\bibinfo{title}{\bibfnamefont{Bell's Inequalities and Quantum Communication Complexity}},
  \bibinfo{journal}{Phys.~Rev.~Lett.} \textbf{\bibinfo{volume}{92}} (\bibinfo{year}{2004}) \bibinfo{pages}{127901}.

\bibitem[{\citenamefont{Collins et~al}(2004)}]{collins04}
\bibinfo{author}{\bibfnamefont{D.} \bibnamefont{Collins}}, \bibnamefont{and}
\bibinfo{author}{\bibfnamefont{N.} \bibnamefont{Gisin}}, 
\bibinfo{title}{\bibfnamefont{A relevant two qubit Bell inequality
inequivalent to the CHSH inequality}},
  \bibinfo{journal}{J. Phys. A: Math. Gen.} \textbf{\bibinfo{volume}{37}} (\bibinfo{year}{2004}) \bibinfo{pages}{1775}.
    
\bibitem[{\citenamefont{Bennett et~al}(1996)}]{bennett96}
\bibinfo{author}{\bibfnamefont{C.~H.} \bibnamefont{Bennett}},
\bibinfo{author}{\bibfnamefont{D.~P.} \bibnamefont{DiVincenzo}}, 
\bibinfo{author}{\bibfnamefont{J.~A.} \bibnamefont{Smolin}}, \bibnamefont{and}
\bibinfo{author}{\bibfnamefont{W.~K.} \bibnamefont{Wootters}}, 
\bibinfo{title}{\bibfnamefont{Mixed-state entanglement and quantum error correction}},
  \bibinfo{journal}{Phys.~Rev.~A} \textbf{\bibinfo{volume}{54}} (\bibinfo{year}{1996}) \bibinfo{pages}{3824}.  
  
\bibitem[{\citenamefont{Wootters et~al}(1998)}]{wootters98}
\bibinfo{author}{\bibfnamefont{W.~K.} \bibnamefont{Wootters}},
\bibinfo{title}{\bibfnamefont{Entanglement of Formation of an Arbitrary State of Two Qubits}},
  \bibinfo{journal}{Phys.~Rev.~Lett.} \textbf{\bibinfo{volume}{80}} (\bibinfo{year}{1998}) \bibinfo{pages}{2245}.
  
\bibitem[{\citenamefont{M\'ethot et al}(2007)}]{methot07}
\bibinfo{author}{\bibfnamefont{A. A.} \bibnamefont{M\'ethot}} \bibnamefont{and}
\bibinfo{author}{\bibfnamefont{V.} \bibnamefont{Scarani}},
\bibinfo{author}{\bibfnamefont{An anomaly of non-locality}},
 \bibinfo{journal}{Quantum Inf. Comput.} \textbf{\bibinfo{volume}{7}} (\bibinfo{year}{2007}) \bibinfo{pages}{157}. 
 
\bibitem[{\citenamefont{Eisert et~al}(1999)}]{eisert99}
\bibinfo{author}{\bibfnamefont{J.} \bibnamefont{Eisert}}, \bibnamefont{and}
\bibinfo{author}{\bibfnamefont{M.~B.} \bibnamefont{Plenio}}, 
\bibinfo{title}{\bibfnamefont{A comparison of entanglement measures}},
  \bibinfo{journal}{J.~Mod.~Opt.} \textbf{\bibinfo{volume}{46}} (\bibinfo{year}{1999}) \bibinfo{pages}{145}. 
 
 \bibitem[{\citenamefont{Dakic et~al}(2010)}]{vedral10}
\bibinfo{author}{\bibfnamefont{B.} \bibnamefont{Daki\'{c}}},
\bibinfo{author}{\bibfnamefont{V.} \bibnamefont{Vedral}}, \bibnamefont{and}
\bibinfo{author}{\bibfnamefont{C.} \bibnamefont{Brukner}}, 
\bibinfo{title}{\bibfnamefont{Necessary and Sufficient Condition for Nonzero Quantum Discord}},
  \bibinfo{journal}{Phys.~Rev.~Lett.} \textbf{\bibinfo{volume}{105}} (\bibinfo{year}{2010}) \bibinfo{pages}{190502}.

\bibitem[{\citenamefont{Costa et~al}(2013)}]{costa13}
\bibinfo{author}{\bibfnamefont{A.~C.~S.} \bibnamefont{Costa}} \bibnamefont{and}
\bibinfo{author}{\bibfnamefont{R.~M.} \bibnamefont{Angelo}}, 
\bibinfo{author}{\bibfnamefont{Bayes' rule, generalized discord, and nonextensive thermodynamics}},
  \bibinfo{journal}{Phys.~Rev.~A} \textbf{\bibinfo{volume}{87}} (\bibinfo{year}{2013}) \bibinfo{pages}{032109}.
  
\bibitem[{\citenamefont{Vidal et~al}(2002)}]{vidal02}
\bibinfo{author}{\bibfnamefont{G.} \bibnamefont{Vidal}}, \bibnamefont{and}
\bibinfo{author}{\bibfnamefont{R.~F.} \bibnamefont{Werner}}, 
\bibinfo{title}{\bibfnamefont{Computable measure of entanglement}},
  \bibinfo{journal}{Phys.~Rev.~A} \textbf{\bibinfo{volume}{65}} (\bibinfo{year}{2002}) \bibinfo{pages}{032314}.

\bibitem[{\citenamefont{Horodecki et~al}(1995)}]{3h95}
\bibinfo{author}{\bibfnamefont{R.} \bibnamefont{Horodecki}},
\bibinfo{author}{\bibfnamefont{P.} \bibnamefont{Horodecki}}, \bibnamefont{and}
\bibinfo{author}{\bibfnamefont{M.} \bibnamefont{Horodecki}}, 
\bibinfo{title}{\bibfnamefont{Violating Bell inequality by mixed spin-1/2 states: necessary and sufficient condition}},
  \bibinfo{journal}{Phys.~Lett.~A} \textbf{\bibinfo{volume}{200}} (\bibinfo{year}{1995}) \bibinfo{pages}{340}.

\bibitem[{\citenamefont{Hu et~al}(2013)}]{hu13}
\bibinfo{author}{\bibfnamefont{M.-L.} \bibnamefont{Hu}}, 
\bibinfo{title}{\bibfnamefont{Relations between entanglement, Bell-inequality violation and teleportation fidelity for the two-qubit X states}},
  \bibinfo{journal}{Quantum Info.~Process.} \textbf{\bibinfo{volume}{12}} (\bibinfo{year}{2013}) \bibinfo{pages}{229}.

\bibitem[{\citenamefont{Yu et~al}(2007)}]{yu07}
\bibinfo{author}{\bibfnamefont{T.} \bibnamefont{Yu}}, \bibnamefont{and}
\bibinfo{author}{\bibfnamefont{J.~H.} \bibnamefont{Eberly}}, 
\bibinfo{title}{\bibfnamefont{Evolution from entanglement to decoherence of bipartite mixed ``X" states}},
  \bibinfo{journal}{Quantum Inf.~Comput.} \textbf{\bibinfo{volume}{7}} (\bibinfo{year}{2007}) \bibinfo{pages}{459}.


\bibitem[{\citenamefont{Bellomo et~al}(2007)}]{bellomo07}
\bibinfo{author}{\bibfnamefont{B.} \bibnamefont{Bellomo}}, \bibinfo{author}{\bibfnamefont{R.} \bibnamefont{Lo Franco}}, \bibnamefont{and}
 \bibinfo{author}{\bibfnamefont{G.} \bibnamefont{Compagno}}, 
\bibinfo{title}{\bibfnamefont{Non-Markovian Effects on the Dynamics of Entanglement}},
  \bibinfo{journal}{Phys.~Rev.~Lett.} \textbf{\bibinfo{volume}{99}} (\bibinfo{year}{2007}) \bibinfo{pages}{160502}.







\end{thebibliography}

\end{document}